\documentclass[final,1p,times]{elsarticle}  % ✅ 正确方式

\usepackage{amssymb}
 \usepackage{amsthm}
\usepackage{amscd}
\usepackage{amsmath}
\usepackage{amsfonts}
\usepackage{amssymb}
\usepackage{color}
\usepackage{graphicx}
\usepackage{url}
\newtheorem{theorem}{Theorem}[section]

\newtheorem{lemma}[subsection]{Lemma}

\usepackage[ruled]{algorithm2e}
\usepackage{mathrsfs}
\usepackage{titletoc}
\usepackage{float}
\usepackage{float}
\usepackage{amsmath}
\usepackage{tikz}
\usepackage{caption}
\usepackage{subcaption}
\usepackage{booktabs}
\usepackage{geometry}
\journal{******}

\begin{document}

\begin{frontmatter}

\title{Finding core subgraphs of directed graphs via discrete Ricci curvature flow}

\author[ruc]{Juan Zhao}
\ead{zhaojuan0509@ruc.edu.cn}

\author[ruc]{Jicheng Ma}
\ead{2019202433@ruc.edu.cn}

\author[ruc]{Yunyan Yang\corref{cor1}}
\ead{yunyanyang@ruc.edu.cn}

\author[bnu]{Liang Zhao}
\ead{liangzhao@bnu.edu.cn}

\cortext[cor1]{Corresponding author}

\address[ruc]{School of Mathematics, Renmin University of China, Beijing, 100872, China}
\address[bnu]{School of Mathematical Sciences, Key Laboratory of Mathematics and Complex Systems of MOE,\\
Beijing Normal University, Beijing, 100875, China}

\begin{abstract}

Ricci curvature and its associated flow offer powerful geometric methods for analyzing complex networks. While existing research heavily focuses on applications for undirected graphs such as community detection and core extraction, there have been relatively less attention on directed graphs.

In this paper, we introduce a definition of Ricci curvature and an accompanying curvature flow for directed graphs. Crucially, for strongly connected directed graphs, this flow admits a unique global solution. We then apply this flow to detect strongly connected subgraphs from weakly connected directed graphs. (A weakly connected graph is connected overall but not necessarily strongly connected).
Unlike prior work requiring graphs to be strongly connected, our method loosens this requirement. We transform a weakly connected graph into a strongly connected one by adding edges with very large artificial weights. This modification does not compromise our core subgraph detection. Due to their extreme weight, these added edges are automatically discarded during the final iteration of the Ricci curvature flow.

For core evaluation, our approach consistently surpasses traditional methods, achieving better results on at least two out of three key metrics. The implementation code is publicly available at https://github.com/12tangze12/Finding-core-subgraphs-on-directed-graphs.

\end{abstract}


\end{frontmatter}

\titlecontents{section}[0mm]
                       {\vspace{.2\baselineskip}}%\bfseries}
                       {\thecontentslabel~\hspace{.5em}}
                        {}
                        {\dotfill\contentspage[{\makebox[0pt][r]{\thecontentspage}}]}
\titlecontents{subsection}[3mm]
                       {\vspace{.2\baselineskip}}%\bfseries}
                       {\thecontentslabel~\hspace{.5em}}
                        {}
                       {\dotfill\contentspage[{\makebox[0pt][r]{\thecontentspage}}]}

\setcounter{tocdepth}{2}
%\tableofcontents

%\setcounter{tocdepth}{1}
%\tableofcontents

%\tableofcontents

\numberwithin{equation}{section}
\section{Introduction}
Curvature serves as a fundamental concept in differential geometry, quantifying the deviation of a manifold from flatness. While the Riemann curvature tensor comprehensively characterizes intrinsic bending, contracting it yields the Ricci curvature tensor. This governs the convergence or divergence of nearby geodesics, with positive Ricci curvature typically promoting convergence. Hamilton introduced Ricci curvature flow \cite{hamilton1982ricci}, an evolution process that smooths a manifold's curvature distribution:
\begin{equation}\label{Hamilton}
\frac{\partial g_{ij}}{\partial t}=-2{\rm Ric}_{ij}.
\end{equation}
This flow proved instrumental in Perelman's resolution of the Poincar\'e conjecture, facilitating the deformation of three-dimensional manifolds toward canonical geometries \cite{perelman2002entropy}.

Discrete analogs of Ricci curvature have been successfully developed for graph structures. Notable formulations include those by Forman \cite{ref17, ref18}, Ollivier \cite{Ollivier1, ref25}, and Lin-Lu-Yau \cite{Lin1}. In particular, Ollivier \cite{ref25} proposed
an analog of (\ref{Hamilton}) for weighted graphs:
\begin{equation}\label{r-flow}
w_e^\prime(t)=-\kappa_e(t)w_e(t),
\end{equation}
where $w_e$ is the edge weight and $\kappa_e$ is the Ricci curvature on edge $e$. In graphs, positive edge curvature indicates a strong relationship between two vertices, whereas negative curvature signals a weak link. Based on this observation and the behavior of the flow in (\ref{Hamilton}),
Ni-Lin-Luo-Gao \cite{Ni1} developed a community detection algorithm using (\ref{r-flow}) combined with topological surgery, while Lai-Bai-Lin \cite{Lai1} employed a normalized Ricci flow based on Lin-Lu-Yau's Ricci curvature to achieve similar results. Regarding the mathematical theory of
(\ref{r-flow}), such as the existence and uniqueness of solutions, it is attributed to Bai-Lin-Lu-Wang-Yau \cite{Bai-Lin}. Recently, Ma-Yang \cite{Ma1,Ma2,Ma3} modified (\ref{r-flow}) into several versions with global solutions. For the convergence of discrete Ricci flow on a weighted graph, we refer readers to Li-M\"unch \cite{Li-Munch}. In addition, Barkanass-Jost-Saucan \cite{Saucan} applied Ricci curvature to network sampling, backbone detection, and structural analysis. Likewise, Zhao-Ma-Yang-Zhao \cite{Zhao} investigated discrete Ricci flows on undirected graphs, deriving bounds on edge weights and demonstrating their effectiveness for core subgraph detection.

Although Ricci curvature and Ricci curvature flow have been extensively studied,  research on their application to directed graphs has been relatively limited. In \cite{Ozawar1}, Ozawa-Sakurai-Yamada extended Lin-Lu-Yau's Ricci curvature to strongly connected directed graphs by employing the mean transition probability kernel associated with the Laplace operator. Eidi-Jost \cite{Eidi} introduced a Ricci curvature for directed hypergraphs as a natural generalization of Ollivier's definition for undirected graphs, based on a carefully designed optimal transport problem between sets of vertices. In \cite{Li}, Li developed another concept of Ricci curvature (flow) for directed graphs and used it to design a community detection algorithm. In \cite{Bai}, Bai-Li-Liu-Lai introduced a rigorous Ricci flow framework for directed weighted graphs, established the existence and uniqueness of its solutions, and demonstrated its effectiveness in capturing structural asymmetry through numerical experiments. More recently, Sengupta-Azarhooshang-Albert-DasGupta \cite{Sengupta} proposed a Ricci curvature flow-based framework for identifying influential cores in both directed and undirected hypergraphs, demonstrating its partial effectiveness on biological and social datasets.

In this paper, we establish Ricci curvature and Ricci curvature flow for directed graphs and study their properties. As an application, we address the problem of core subgraph detection. According to \cite{2011,1999,1983,Sengupta}, a core subgraph is a tightly connected subgraph whose removal significantly alters the topology of the entire graph. Core subgraphs are generally not unique and require specific identification methods. Here we use a discrete Ricci curvature flow combined with an edge deletion strategy to extract core subgraphs from directed graphs. The process consists of four steps:
$(i)$ Weakly connected directed graphs are transformed into strongly connected graphs by adding artificial edges, enabling Ricci curvature to be defined on all edges; $(ii)$ The discrete Ricci curvature flow is run for finitely many iterations; $(iii)$ After the final iteration, all artificial edges and real edges with the largest weights are deleted, along with their incident nodes; $(iv)$ The remaining nodes induce a subgraph, and its largest strongly connected component is selected as the core subgraph. Experimental results on real-world networks show that our algorithm outperforms classical methods, including Pagerank, degree centrality, betweenness centrality, and closeness centrality, achieving superior results on at least two structural metrics.

The remainder of the paper is organized as follows. Section 2 introduces notations and main results. Section 3 presents proofs of the main results. Section 4 defines three core subgraph metrics, provides illustrative examples, and presents the algorithm. Section 5 applies the algorithm to real-world networks for performance assessment. Section 6 provides concluding remarks.

\section{Notations and main results}

\begin{figure}[H]
\centering
\begin{tikzpicture}[scale=1, >=stealth]

% ---------------- G1: Strongly connected ----------------
\begin{scope}[xshift=0cm]
  \coordinate (x) at (0,1);
  \coordinate (y) at (-1,-0.5);
  \coordinate (z) at (1,-0.5);

  \draw[->, thick] (x) -- (y) node[midway,left]  {\small 1};
  \draw[->, thick] (y) -- (z) node[midway,below] {\small 1};
  \draw[->, thick] (z) -- (x) node[midway,right] {\small 1};

  \fill (x) circle (1.5pt) node[above=3pt] {\small $x$};
  \fill (y) circle (1.5pt) node[left=3pt]  {\small $y$};
  \fill (z) circle (1.5pt) node[right=3pt] {\small $z$};

  \node[below=12pt] at (0,-0.8) {\scriptsize Strongly connected $G_1$};
\end{scope}

% ---------------- G2: Weakly connected ----------------
\begin{scope}[xshift=5.5cm]
 
  \coordinate (x) at (0,1);
  \coordinate (y) at (-1,-0.5);
  \coordinate (z) at (1,-0.5);

  \draw[->, thick] (x) -- (y) node[midway,left]  {\small 1};
  \draw[->, thick] (y) -- (z) node[midway,below] {\small 1};
  \draw[->, thick] (x) -- (z) node[midway,right] {\small 1};

  \fill (x) circle (1.5pt) node[above=3pt] {\small $x$};
  \fill (y) circle (1.5pt) node[left=3pt]  {\small $y$};
  \fill (z) circle (1.5pt) node[right=3pt] {\small $z$};

  \node[below=12pt] at (0,-0.8) {\scriptsize Weakly connected $G_2$};
\end{scope}

\end{tikzpicture}
\caption{Illustration of Example 1.}
\label{figer1} 
\end{figure}
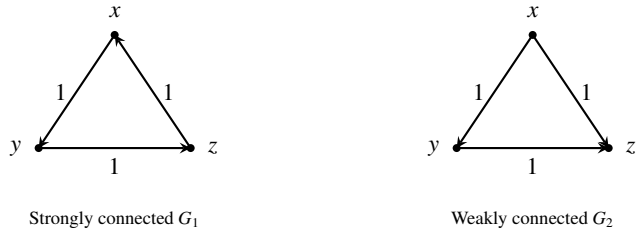

Let $G=(V,E,\mathbf{w})$ be a directed graph, where $V=\{z_1,z_2,\dots,z_n\}$ is the vertex set, $E=\{e_1,e_2,\dots,e_m\}$ is the edge set of directed edges, and $\mathbf{w}=(w_{e_1},w_{e_2},\dots,w_{e_m})\in\mathbb{R}^m_+$ is a vector of edge weights. To simplify notation, we denote a directed edge $e \in E$ from $x$ to $y$ as $e=xy$. Let us recall two concepts on $G$:
 \begin{itemize}
  \item $G$ is \textit{weakly connected} if for any $u,v\in V$, there exist an integer $k$ and a vertex sequence $u=x_0, x_1,\dots,x_k=v$ such that $x_{i-1} \sim x_i$ for all $i=1,\dots,k$, where $x_{i-1} \sim x_i$ means $x_{i-1}$ and $x_i$ are adjacent (i.e., either $x_{i-1}x_i \in E$ or $x_i x_{i-1} \in E$).
  \item $G$ is \textit{strongly connected} if for any $u,v\in V$, there exist directed paths $u \to v$ and $v \to u$. Specifically, there exist integers $p,q$ and vertex sequences $\{y_i\}_{i=0}^p$, $\{z_j\}_{j=0}^q$ such that
      $y_0=u$, $y_p=v$, and $y_{i-1}y_i \in E$ for $i=1,\dots,p$ ($u\to v$ path);
      $z_0=v$, $z_q=u$, and $z_{j-1}z_j \in E$ for $j=1,\dots,q$ ($v\to u$ path).
 \end{itemize}
 {\it Example 1}. Consider the vertex set $V=\{x,y,z\}$ with edge weights $w_{xy}=w_{yz}=w_{zx}=w_{xz}=1$. Define
  \begin{itemize}
  \item $G_1=(V,E_1,\textbf{w}_1)$, where $E_1=\{xy,yz,zx\}$ and $\textbf{w}_1=(w_{xy},w_{yz},w_{zx})$
  \item $G_2=(V,E_2,\textbf{w}_2)$, where $E_2=\{xy,yz,xz\}$ and $\textbf{w}_2=(w_{xy},w_{yz},w_{xz})$
  \end{itemize}
Figure~\ref{figer1} illustrates that $G_1$ is strongly connected, while $G_2$ is weakly connected.

The distance from $z_i$ to $z_j$ is defined as
\begin{equation*}d(z_i,z_j)=\left\{\begin{array}{lll}\inf_{\gamma}\sum_{e\in\gamma}w_e,&{\rm if\,\,there\,\, is\,\, a\,\,path}\,\,\gamma\,\,{\rm from}\,\,z_i\,\,{\rm to}\,\,z_j\\[1.5ex]+\infty,&{\rm if\,\,there\,\, is\,\, no\,\,path\,\, from}\,\,z_i\,\,{\rm to}\,\,z_j,\end{array}\right.
\end{equation*}
 where the infimum is taken over all paths $\gamma$ from $z_i$ to $z_j$.
  For any $x\in V$, its outward neighbor set is denoted  by
 $$\mathscr{N}_x^{\rm out}=\{u\in V: xu\in E\}.$$
 Given $\alpha\in [0,1]$,
 an outward $\alpha$-lazy one-step random walk at $x$ is defined as
\begin{equation*}\mu_{x,\alpha}^{\rm out}(z)=\left\{\begin{array}{lll}
\alpha&{\rm if}& z=x\\[1.2ex]
(1-\alpha)\frac{w_{xz}}{\sum_{xu\in E}w_{xu}}&{\rm if}& xz\in E\\[1.2ex]
0&{\rm if}&z\in V\setminus (\{x\}\cup\mathscr{N}_x^{\rm out}).
\end{array}\right.\end{equation*}
Clearly, $\mu_{x,\alpha}^{\rm out}$ is a probability measure on $V$.  For any two
probability measures $\mu_1$ and $\mu_2$,
the Wasserstein distance from $\mu_1$ to $\mu_2$ is given by
$$W(\mu_1,\mu_2)=\inf_A\sum_{u,v\in V}A(u,v)d(u,v),$$
where $A$ is taken over all couplings between  $\mu_1$ and $\mu_2$. Motivated by \cite{Lin1,Ollivier1,Ozawar1}, for any $\alpha\in[0,1]$ and directed edge $e=xy$, we define the
Ricci curvature on $e$ as
\begin{equation}\label{Olli-Ric}
\kappa_{e}^\alpha=1-\frac{W(\mu_{x,\alpha}^{\rm out},\mu_{y,\alpha}^{\rm out})}{d(x,y)}.
\end{equation}
Similar to \cite{Ma1,Ma2}, the Ricci curvature flow associated with $\kappa_e^\alpha$ reads
\begin{equation}\label{ric-flow}
\left\{\begin{array}{lll}
w_{e}^\prime(t)=-\kappa_{e}^\alpha(t) \rho_{e}(t)\\[1.2ex]
w_{e}(t)>0,\,w_{e}(0)=w_{0,e}\\[1.2ex]
\forall e\in E.
\end{array}\right.
\end{equation}

Our first result is the following:

\begin{theorem}\label{thm1}
Assume $G=(V,E,\mathbf{w}_0)$ is a strongly connected directed graph, where $\mathbf{w}_0=(w_{0,e})_{e\in E}$ is the initial weights
on $E$. Let $\alpha$ be a number in $[0,1]$, $\kappa_{e}^\alpha$ be the Ricci curvature
defined as in (\ref{Olli-Ric}), and $\rho_e=d(x,y)$ denotes the length of a directed edge $e=xy$. Then the Ricci curvature flow (\ref{ric-flow})
has a unique global solution $\mathbf{w}(t)=(w_{e_1}(t),\cdots,w_{e_m}(t))$ for $t\in[0,+\infty)$.
\end{theorem}

For simplicity, we denote $\kappa_e=\kappa_e^\alpha$ for any fixed $\alpha\in[0,1]$. Observing that (\ref{ric-flow}) is a continuous Ricci curvature flow, we write a discrete version of (\ref{ric-flow}) as
\begin{equation}\label{disc-flow}
\left\{\begin{array}{lll}
w_e^{(j+1)}=w_e^{(j)}-s\kappa_e^{(j)}\rho_e^{(j)}\\[1.2ex]
w_e^{(j)}>0,\,\,\forall j\in\mathbb{N}\\[1.2ex]
w_e^{(0)}=w_{0,e},\,\,\forall e\in E.
\end{array}\right.
\end{equation}

As an analog of \cite{Zhao}, our second result is stated as follows:
\begin{theorem}\label{discrete-thm}
Under the same assumptions as in Theorem \ref{thm1}, if $0<s<1$, then the discrete Ricci curvature flow (\ref{disc-flow})
has a unique solution $(w_{e}^{(j)})$ for all $e\in E$ and all $j\in\mathbb{N}$. In particular, there holds
$$(1-s)^jw_{0,e}\leq w_e^{(j)}\leq (1+ms)^j\sum_{\tau\in E}w_{0,\tau}$$
for all $e\in E$ and all $j\in\mathbb{N}$.
\end{theorem}

\section{Proof of Theorems \ref{thm1} and \ref{discrete-thm}}

In this section, we shall prove Theorems \ref{thm1} and \ref{discrete-thm}. Let us first prove that both distance and Wasserstein distance
are all Lipschitz in the weights $\textbf{w}$. 

 \begin{lemma}\label{distance}
 If $G=(V,E,\mathbf{w})$ is a strongly connected directed graph, then for any two vertices $x$ and $y$, $d(x,y)$ is Lipschitz with respect to
 $\mathbf{w}$.
 \end{lemma}

 \proof Fix $x$ and $y$. Given any two vectors $\mathbf{w}^{(1)}=(w_{e_1}^{(1)},\cdots,w_{e_m}^{(1)})$,
 $\mathbf{w}^{(2)}=(w_{e_1}^{(2)},\cdots,w_{e_m}^{(2)})\in\mathbb{R}^m_+$,
 we denote the directed distance from $x$ to $y$ with respect to $\mathbf{w}=\mathbf{w}^{(1)}$ or $\mathbf{w}=\mathbf{w}^{(2)}$
 by $$d^{(1)}(x,y)=\inf_{\gamma}\sum_{e\in \gamma}w_e^{(1)}$$
 and $${d}^{(2)}(x,y)=\inf_{\gamma}\sum_{e\in \gamma}{w}_e^{(2)}$$
 respectively, where $\gamma$ is taken over all directed paths from $x$ to $y$.

 We now distinguish two cases to proceed.

 {\it Case 1.} $d^{(1)}(x,y)\leq d^{(2)}(x,y)$.

 Since there exists a directed path $\gamma_1$ from $x$ to $y$ such that
 $d^{(1)}(x,y)=\sum_{e\in \gamma_1}w^{(1)}_e$, we have
 \begin{eqnarray*}
 0&\leq& d^{(2)}(x,y)-\sum_{e\in \gamma_1}w^{(1)}_e
 \,\leq\,\sum_{e\in \gamma_1}w^{(2)}_e-\sum_{e\in \gamma_1}w^{(1)}_e\\
 &\leq&\sum_{e\in \gamma_1}|w^{(2)}_e-w^{(1)}_e|
 \,\leq\, \sqrt{m}|\mathbf{w}^{(2)}-\mathbf{w}^{(1)}|.
 \end{eqnarray*}
Here, $|\mathbf{w}^{(2)}-\mathbf{w}^{(1)}| = \sqrt{\sum_{i=1}^{m} (w_{e_{i}}^{(2)}-w_{e_{i}}^{(1)})^2}$.

 {\it Case 2.} $d^{(1)}(x,y)> d^{(2)}(x,y)$.

 Take a directed path $\gamma_2$ from $x$ to $y$ such that
$d^{(2)}(x,y)=\sum_{e\in \gamma_2}w^{(2)}_e$. It follows that
\begin{eqnarray*}
 0&<& d^{(1)}(x,y)-\sum_{e\in \gamma_2}w^{(2)}_e
 \,\leq\,\sum_{e\in \gamma_2}w^{(1)}_e-\sum_{e\in \gamma_2}w^{(2)}_e\\
 &\leq&\sum_{e\in \gamma_2}|w^{(1)}_e-w^{(2)}_e|
 \,\leq\, \sqrt{m}|\mathbf{w}^{(1)}-\mathbf{w}^{(2)}|.
 \end{eqnarray*}
Combining Cases 1 and 2, we conclude
\begin{equation}\label{dist-lip}|d^{(1)}(x,y)-d^{(2)}(x,y)|\leq \sqrt{m}|\mathbf{w}^{(1)}-\mathbf{w}^{(2)}|,
\end{equation}
which gives the desired result. $\hfill\Box$\\

\begin{lemma}\label{W-distance}
 If $G=(V,E,\mathbf{w})$ is a strongly connected directed graph, then for any $xy\in E$, the directed Wasserstein distance
 $W(\mu_{x,\alpha}^{\rm out},\mu_{y,\alpha}^{\rm out})$ is Lipschitz with respect to
 $\mathbf{w}$.
 \end{lemma}

 \proof Fix $e=xy\in E$. For any two vectors $\mathbf{w}^{(1)}=(w_{e_1}^{(1)},\cdots,w_{e_m}^{(1)})$ and $\mathbf{w}^{(2)}=
 (w_{e_1}^{(2)},\cdots,w_{e_m}^{(2)})\in\mathbb{R}^m_+$ satisfying
 \begin{equation}\label{hypo}\Lambda^{-1}\leq w_{e}^{(1)}\leq\Lambda,\,\,
\Lambda^{-1}\leq {w}_{e}^{(2)}\leq\Lambda,\,\,
|w_{e}^{(1)}-{w}_{e}^{(2)}|\leq \delta \end{equation}
for two positive constants $\Lambda$ and $\delta$,
 we denote the directed distances from
 $x$ to $y$  and the directed Wasserstein distances from $\mu_{x,\alpha}^{\rm out}$ to $\mu_{y,\alpha}^{\rm out}$ by $d^{(1)}(x,y)$,
 $W^{(1)}(\mu_{x,\alpha}^{\rm out},\mu_{y,\alpha}^{\rm out})$ with respect to $\mathbf{w}=\mathbf{w}^{(1)}$, and $d^{(2)}(x,y)$,
 $W^{(2)}(\mu_{x,\alpha}^{\rm out},\mu_{y,\alpha}^{\rm out})$ with respect to $\mathbf{w}=\mathbf{w}^{(2)}$, respectively.

 By the Kantorovich-Rubinstein duality formula,
\begin{eqnarray}\label{W1}
W^{(1)}(\mu_{x,\alpha}^{\rm out},\mu_{y,\alpha}^{\rm out})=\sup_{\psi\in{\rm Lip}^{(1)}1}\sum_{u\in V} \psi(u)(\mu_{x,\alpha}^{\rm out}(u)-\mu_{y,\alpha}^{\rm out}(u)),\\
W^{(2)}(\mu_{x,\alpha}^{\rm out},\mu_{y,\alpha}^{\rm out})=\sup_{\psi\in{\rm Lip}^{(2)}1}\sum_{u\in V} \psi(u)(\mu_{x,\alpha}^{\rm out}(u)-\mu_{y,\alpha}^{\rm out}(u)),\label{W2}
\end{eqnarray}
where
\begin{eqnarray*}
&&{\rm Lip}^{(1)}1=\left\{f\in V^{\mathbb{R}}: f(u)-f(v)\leq d^{(1)}(u,v),\,\forall u,v\in V\right\},\\
&&{\rm Lip}^{(2)}1=\{f\in V^{\mathbb{R}}: f(u)-f(v)\leq d^{(2)}(u,v),\,\forall u,v\in V\}.
\end{eqnarray*}
Let $p\in V$, $i\in\{1,2\}$  and $\psi\in {\rm Lip}^{(i)}1$. Set $\tilde{\psi}(u)=\psi(u)-\psi(p)$ for all $u\in V$.
Then $\tilde{\psi}\in {\rm Lip}^{(i)}1$  and
\begin{equation}\label{equa}\sum_{u\in V} \psi(u)(\mu_{x,\alpha}^{\rm out}(u)-\mu_{y,\alpha}^{\rm out}(u))=\sum_{u\in V} \tilde{\psi}(u)(\mu_{x,\alpha}^{\rm out}(u)-\mu_{y,\alpha}^{\rm out}(u))\end{equation}
with respect to $\mathbf{w}=\mathbf{w}^{(i)}$. Since
\begin{eqnarray*}
\tilde{\psi}(u)=\psi(u)-\psi(p)\leq d^{(i)}(u,p),\,\, -\tilde{\psi}(u)=\psi(p)-\psi(u)\leq d^{(i)}(p,u),
\end{eqnarray*}
there holds $|\tilde{\psi}(u)|\leq D^{(i)}:=\max_{v,s\in V}d^{(i)}(v,s)$ for all $u\in V$. This together with (\ref{W1}), (\ref{W2}) and
(\ref{equa}) leads to
\begin{equation}\label{W}
W^{(i)}(\mu_{x,\alpha}^{\rm out},\mu_{y,\alpha}^{\rm out})=\sup_{\psi\in{\rm Lip}^{(i)}1,\,|\psi|\leq D^{(i)}}\sum_{u\in V} \psi(u)(\mu_{x,\alpha}^{\rm out}(u)-\mu_{y,\alpha}^{\rm out}(u))
\end{equation}
for $i=1,2$. We distinguish two cases to proceed.\\

{\it Case 1.} $W^{(1)}(\mu_{x,\alpha}^{\rm out},\mu_{y,\alpha}^{\rm out})\leq W^{(2)}(\mu_{x,\alpha}^{\rm out},\mu_{y,\alpha}^{\rm out})$.

By (\ref{dist-lip}) in the proof of Lemma \ref{distance},
$$|d^{(1)}(u,v)-d^{(2)}(u,v)|\leq \sqrt{m}|\mathbf{w}^{(1)}-\mathbf{w}^{(2)}|,\quad \forall u,v\in V.$$
In view of (\ref{hypo}), we have
$$\Lambda^{-1}\leq d^{(1)}(u,v)\leq m\Lambda,\,\,\Lambda^{-1}\leq d^{(2)}(u,v)\leq m\Lambda,\quad \forall u,v\in V.$$
It follows that
\begin{equation}\label{comp}
\frac{d^{(2)}(u,v)}{d^{(1)}(u,v)}=1+\frac{d^{(2)}(u,v)-{d}^{(1)}(u,v)}{{d}^{(1)}(u,v)}
\leq1+\sqrt{m}\Lambda|\mathbf{w}^{(1)}-{\mathbf{w}^{(2)}}|.
\end{equation}
In view of (\ref{W}), there exists some $\psi\in {\rm Lip}^{(2)}1$ with $|\psi|\leq D^{(2)}$ on $V$ such that
$$W^{(2)}(\mu_{x,\alpha}^{\rm out},\mu_{y,\alpha}^{\rm out})=\sum_{u\in V} \psi(u)(\mu_{x,\alpha}^{\rm out}(u)-\mu_{y,\alpha}^{\rm out}(u)).$$
Set
$$\widetilde{\psi}(u)=\frac{\psi(u)}{1+\sqrt{m}\Lambda|\mathbf{w}^{(1)}-\mathbf{w}^{(2)}|},\quad\forall u\in V.$$
Using the relation (\ref{comp}), we have
$$\widetilde{\psi}(u)-\widetilde{\psi}(v)=\frac{\psi(u)-\psi(v)}{1+\sqrt{m}\Lambda|\mathbf{w}^{(1)}-\mathbf{w}^{(2)}|}
\leq\frac{d^{(2)}(u,v)}{1+\sqrt{m}\Lambda|\mathbf{w}^{(1)}-\mathbf{w}^{(2)}|}\leq {d}^{(1)}(u,v)$$
for all $u,v\in V$, and thus $\widetilde{\psi}\in {\rm Lip}^{(1)}1$.
We rewrite $\mu_{x,\alpha,\mathbf{w}}^{\rm out}$ and $\mu_{y,\alpha,\mathbf{w}}^{\rm out}$ instead of $\mu_{x,\alpha}^{\rm out}$ and
$\mu_{y,\alpha}^{\rm out}$ with respect to $\mathbf{w}$ respectively. Clearly, there exists a constant $C_1$ depending only on $\Lambda$ and $\delta$ such that
$$|\mu_{y,\alpha,\mathbf{w}^{(1)}}^{\rm out}(u)-\mu_{y,\alpha,\mathbf{w}^{(2)}}^{\rm out}(u)|\leq C_1|\mathbf{w}^{(1)}-\mathbf{w}^{(2)}|$$
for all $u\in V$. We calculate
\begin{eqnarray*}
0&\leq& W^{(2)}(\mu_{x,\alpha}^{\rm out},\mu_{y,\alpha}^{\rm out})-W^{(1)}(\mu_{x,\alpha}^{\rm out},\mu_{y,\alpha}^{\rm out})\\&=&
\sum_{u\in V}{\psi}(u)(\mu_{x,\alpha,\mathbf{w}^{(2)}}^{\rm out}(u)-\mu_{y,\alpha,\mathbf{w}^{(2)}}^{\rm out}(u))-W^{(1)}(\mu_{x,\alpha}^{\rm out},\mu_{y,\alpha}^{\rm out})\\
&\leq&\sum_{u\in V}{\psi}(u)(\mu_{x,\alpha,\mathbf{w}^{(2)}}^{\rm out}(u)-\mu_{y,\alpha,\mathbf{w}^{(2)}}^{\rm out}(u))-
\sum_{u\in V}\widetilde{\psi}(u)(\mu_{x,\alpha,\mathbf{w}^{(1)}}^{\rm out}(u)-\mu_{y,\alpha,\mathbf{w}^{(1)}}^{\rm out}(u))\\[1.2ex]
&\leq& \sum_{u\in V}|\psi(u)|(|\mu_{x,\alpha,\mathbf{w}^{(2)}}^{\rm out}(u)-\mu_{x,\alpha,\mathbf{w}^{(1)}}^{\rm out}(u)|+|\mu_{y,\alpha,\mathbf{w}^{(2)}}^{\rm out}(u)-\mu_{y,\alpha,\mathbf{w}^{(1)}}^{\rm out}(u)|)\\
&&+\sum_{u\in V} |\psi(u)-\widetilde{\psi}(u)|(\mu_{x,\alpha,\mathbf{w}^{(1)}}^{\rm out}(u)+\mu_{y,\alpha,\mathbf{w}^{(1)}}^{\rm out}(u))\\[1.2ex]
&\leq& 2n(C_1+\sqrt{m}\Lambda)D^{(2)}|\mathbf{w}^{(1)}-\mathbf{w}^{(2)}|,
\end{eqnarray*}
where $n$ is the total number of vertices of $V$. Moreover, it follows from (\ref{hypo}) that $D^{(2)}\leq C$ for some constant $C$
depending only on $m$, $\Lambda$ and $\delta$. \\

{\it Case 2.} $W^{(1)}(\mu_{x,\alpha}^{\rm out},\mu_{y,\alpha}^{\rm out})> W^{(2)}(\mu_{x,\alpha}^{\rm out},\mu_{y,\alpha}^{\rm out})$.

Using the same argument as in Case 1, we obtain
$$0\leq W^{(1)}(\mu_{x,\alpha}^{\rm out},\mu_{y,\alpha}^{\rm out})- W^{(2)}(\mu_{x,\alpha}^{\rm out},\mu_{y,\alpha}^{\rm out})\leq C_2
|\mathbf{w}^{(1)}-\mathbf{w}^{(2)}|$$ for some constant $C_2$ depending only on $n$, $m$, $\Lambda$, $\delta$.

Combining the above two cases, we complete the proof of the lemma. $\hfill\Box$\\

{\it Proof of Theorem \ref{thm1}.}

 We may write $\mathbf{w}=(w_{e_1},\cdots,w_{e_m})\in\mathbb{R}^m_+$ and $\mathbf{F}(\mathbf{w})=
(-\kappa_{e_1}\rho_{e_1},\cdots,-\kappa_{e_m}\rho_{e_m})\in\mathbb{R}^m_+$, since $\kappa_e$ and $\rho_e$ are uniquely determined by the weight
$\mathbf{w}$. It follows from Lemmas \ref{dist-lip} and \ref{W-distance} that $\mathbf{F}(\mathbf{w})$ is locally Lipschitz with respect to $\mathbf{w}$ in $\mathbf{R}^m_+$. Hence there exists some $T>0$ such that the ordinary system
\begin{equation*}
\left\{
\begin{array}{lll}
\mathbf{w}^\prime(t)=\mathbf{F}(\mathbf{w}(t))\\[1.2ex]
\mathbf{w}(t)\in\mathbb{R}^m_+,\,
\mathbf{w}(0)=\mathbf{w}_0\in\mathbb{R}^m_+
\end{array}\right.
\end{equation*}
has a unique local solution $\mathbf{w}(t)$ for $t\in[0,T)$. Since
$$-\kappa_e\rho_e=W(\mu_{x,\alpha}^{\rm out},\mu_{y,\alpha}^{\rm out})-d(x,y),$$
we easily get
$$-w_e\leq -\kappa_e\rho_e\leq \sum_{\tau\in E}w_\tau,$$
or equivalently
\begin{equation*}
-w_e(t)\leq w_e^\prime(t)\leq \sum_{\tau\in E}w_\tau(t),\quad\forall t\in[0,T).
\end{equation*}
This leads to
\begin{equation*}
w_e(0)\exp(-t)\leq w_e(t)\leq w_e(0)\exp(mt),\quad\forall t\in[0,T).
\end{equation*}
Therefore, by the ODE theory (\cite{ODE}, Chapter 6), $\mathbf{w}(t)$ can be extended to $t\in [0,+\infty)$. $\hfill\Box$\\

{\it Proof of Theorem \ref{discrete-thm}.}

Fix $\alpha\in[0,1)$, $s\in(0,1)$, $t_j=js$ for any $j\in\mathbb{N}$ and $e=xy\in E$. Denote the Wasserstein distance
and the Ricci curvature at $t_j$ by $W^{(j)}$ and $\kappa^{(j)}$ respectively. We estimate
$$W^{(j)}(\mu_{x,\alpha}^{\rm out},\mu_{y,\alpha}^{\rm out})\leq \sum_{u,v\in V}A(u,v)\rho^{(j)}(u,v)\leq \sum_{u,v\in V}A(u,v)\sum_{\tau\in E} w_\tau^{(j)}
=\sum_{\tau\in E} w_\tau^{(j)},$$
where $A:V\times V\rightarrow[0,1]$ is any coupling between $\mu_{x,\alpha}^{\rm out}$ and $\mu_{y,\alpha}^{\rm out}$, i.e., $\sum_{v\in V}A(u,v)=\mu_{x,\alpha}^{\rm out}(u)$, $\sum_{u\in V}A(u,v)=\mu_{y,\alpha}^{\rm out}(v)$ and $\sum_{u,v\in V}A(u,v)=1$. It follows that
$$\kappa_e^{(j)}=1-\frac{W^{(j)}(\mu_{x,\alpha}^{\rm out},\mu_{y,\alpha}^{\rm out})}{\rho_e^{(j)}}\geq 1-\frac{\sum_{\tau\in E} w_\tau^{(j)}}{\rho_e^{(j)}},$$
which together with $\kappa_e^{(j)}\leq 1$ leads to
\begin{equation}\label{right}-\sum_{\tau\in E} w_\tau^{(j)}\leq \kappa_e^{(j)}\rho_e^{(j)}\leq w_e^{(j)}.\end{equation}
In view of (\ref{right}),  we obtain by (\ref{disc-flow}),
\begin{equation*}
(1-s)w_e^{(j)}\leq w_e^{(j+1)}\leq w_e^{(j)}+s\sum_{\tau\in E} w_\tau^{(j)}.
\end{equation*}
This immediately gives
\begin{equation*}
(1-s)w_e^{(j)}\leq w_e^{(j+1)}\leq \sum_{\tau\in E}w_\tau^{(j+1)}\leq (1+ms)\sum_{\tau\in E}w_\tau^{(j)}.
\end{equation*}
By an induction argument, we have
\begin{equation*}
(1-s)^{j+1}w_{0,e}\leq w_e^{(j+1)}\leq (1+ms)^{j+1}\sum_{e\in E}w_{0,e},
\end{equation*}
which gives the desired result. $\hfill\Box$

\section{Core subgraphs and Ricci curvature flow}
In this section, we first introduce three metrics for evaluating core subgraphs. To ensure Ricci curvature flow is well defined, the original graph must be strongly connected. For weakly connected graphs, we employ standard edge augmentation procedure from graph theory \cite{Eswaran1976}, which adds the minimal number of edges to achieve strong connectivity. We then illustrate our core detection algorithm, which combines discrete Ricci curvature flow (\ref{disc-flow}) with an edge deletion strategy, through two examples. Finally, we present the corresponding algorithm.

\subsection{Core subgraphs and their metrics}
Several studies have proposed structural metrics to evaluate core subgraphs in undirected graphs \cite{Albert, Kitazono, Koujaku}, yet analogous measures for directed networks remain scarce. Following the framework for finding influential cores via normalized Ricci curvature flows in undirected and directed hypergraphs \cite{Sengupta}, we employ three structural indicators to quantify the cohesiveness and influence of the core subgraphs extracted using our Ricci curvature flow method.

Let $G=(V,E)$ be a directed graph.
A directed subgraph $G^\prime=(V^\prime,E^\prime)$ is called a {\it core subgraph} if $V^\prime\subset V$,
$E^\prime\subset E$, and $G^\prime$ is strongly connected. A core subgraph represents a tightly connected structure whose removal substantially changes the overall topology of the graph.
Denote the induced subgraph of $V\setminus V^\prime$ by $G^\ast$, and let $\xi$ be the number of node pairs $\{u,v\}\subseteq V \setminus V^{\prime}$ that remain connected in $G^\ast$.
We define three metrics to evaluate $G^{\prime}$ as follows:
\begin{equation*}
r_{d}^{\mathrm{in}}=\frac{1}{|V^\prime|}\sum_{x\in V^\prime}\frac{\deg^{\mathrm{in}}_{G^\prime}(x)}{\deg^{\mathrm{in}}_G(x)}, \quad
r_{d}^{\mathrm{out}}=\frac{1}{|V^\prime|}\sum_{x\in V^\prime}\frac{\deg^{\mathrm{out}}_{G^\prime}(x)}{\deg^{\mathrm{out}}_G(x)},
\end{equation*}
and
\begin{equation*}
r_{s}=\frac{1}{\xi} \sum_{\{u,v\}\subset V \setminus V^{\prime},\, \text{dist}_{G^\ast}(u,v) < \infty} \frac{\text{dist}_{G^\ast}(u,v)}{\text{dist}_G(u,v)}.
\end{equation*}
Here, $\deg^{\mathrm{in}}_{G^\prime}(x)$ and $\deg^{\mathrm{out}}_{G^\prime}(x)$ represent the numbers of incoming and outgoing edges of node $x$ in the core subgraph $G^\prime$, $\deg^{\mathrm{in}}_G(x)$ and $\deg^{\mathrm{out}}_G(x)$ are the corresponding numbers in the original graph $G$.
$|V^\prime|$ is the number of nodes in $V^\prime$. $\text{dist}_{G}$ and $\text{dist}_{G^\ast}$ are directed graph distances on $G$ and $G^\ast$ (each edge has length $1$). By definition, $0\leq r_{d}^{\mathrm{in}}, r_{d}^{\mathrm{out}}\leq 1$, $r_s\geq 1$.
In general, as $r_{d}^{\mathrm{in}}$ and $r_{d}^{\mathrm{out}}$ approach $1$, the connections in the core subgraph become tighter; whereas a larger $r_s$ indicates that shortest paths between node pairs in the residual subgraph $G^\ast$ are more likely to pass through the core nodes.

\subsection{From weak connectivity to strong connectivity}
In many real-world scenarios, a directed graph is not strongly connected.  For a connected but not strongly connected directed graph $G$, there exist edges $e=xy$ such that no directed path exists from $y$ to $x$. In theory and application, the distance from $y$ to $x$ is assumed to be infinite. This leads to invalidity of the Wasserstein distance $W(\mu_{x,\alpha}^{\rm out},\mu_{y,\alpha}^{\rm out})$ and thus invalidity of Ricci curvature on $e$. 

To address this issue, we transform a weakly connected graph into a strongly connected graph by adding a minimal set of edges, following a standard procedure in graph theory. Specifically, the graph is first decomposed into some strongly connected components, and a condensed graph is constructed in which each component is represented by a node. Source components (with no incoming edges) and sink components (with no outgoing edges) are then connected iteratively via representative nodes until the graph becomes strongly connected. Each added edge is assigned a sufficiently large weight $A>0$ and marked as artificial. The resulting graph, denoted by $G_A$, ensures that the Wasserstein distance and Ricci curvature are well defined on all edges.
The following example illustrates how to add the minimal edges to achieve strong connectivity.\\

\begin{figure}[H]
\centering
\begin{tikzpicture}[scale=0.6, >=stealth]

% ---------------- G: Not strongly connected ----------------
\begin{scope}[xshift=0cm]
  % 顶点坐标
  \coordinate (x) at (0,0);
  \coordinate (y) at (2.5,0);
  \coordinate (z1) at (-1.5,1.5);
  \coordinate (z2) at (-1.5,-1.5);
  \coordinate (z3) at (4,1.5);
  \coordinate (z4) at (4,-1.5);

  % 有向边和权重
  \draw[->, thick] (x) -- (y) node[midway, above] {\small 1};
  \draw[->, thick] (z1) -- (x) node[midway, left] {\small 1};
  \draw[->, thick] (z2) -- (x) node[midway, left] {\small 1};
  \draw[->, thick] (y) -- (z3) node[midway, above] {\small 1};
  \draw[->, thick] (y) -- (z4) node[midway, below] {\small 1};

  % 顶点
  \foreach \p in {x,y,z1,z2,z3,z4} { \fill (\p) circle (1.5pt); }
  \node[above=2pt] at (x) {\small $x$};
  \node[above=2pt] at (y) {\small $y$};
  \node[left=2pt]  at (z1) {\small $z_1$};
  \node[left=2pt]  at (z2) {\small $z_2$};
  \node[right=2pt] at (z3) {\small $z_3$};
  \node[right=2pt] at (z4) {\small $z_4$};

  \node[below=8pt] at (1,-2) {\scriptsize Original graph $G$};
\end{scope}

% ---------------- G_A: Strongly connected ----------------
\begin{scope}[xshift=9cm]
  \coordinate (x) at (0,0);
  \coordinate (y) at (2.5,0);
  \coordinate (z1) at (-1.5,1.5);
  \coordinate (z2) at (-1.5,-1.5);
  \coordinate (z3) at (4,1.5);
  \coordinate (z4) at (4,-1.5);

  \draw[->, thick] (x) -- (y) node[midway, above] {\small 1};
  \draw[->, thick] (z1) -- (x) node[midway, left] {\small 1};
  \draw[->, thick] (z2) -- (x) node[midway, left] {\small 1};
  \draw[->, thick] (y) -- (z3) node[midway, above] {\small 1};
  \draw[->, thick] (y) -- (z4) node[midway, below] {\small 1};

  \draw[->, thick, red, dashed] (z3) -- (z1) node[midway, above] {\small $A$};
  \draw[->, thick, red, dashed] (z4) -- (z2) node[midway, below] {\small $A$};

  \foreach \p in {x,y,z1,z2,z3,z4} { \fill (\p) circle (1.5pt); }
  \node[above=2pt] at (x) {\small $x$};
  \node[above=2pt] at (y) {\small $y$};
  \node[left=2pt]  at (z1) {\small $z_1$};
  \node[left=2pt]  at (z2) {\small $z_2$};
  \node[right=2pt] at (z3) {\small $z_3$};
  \node[right=2pt] at (z4) {\small $z_4$};

  \node[below=8pt] at (1,-2) {\scriptsize Strongly connected graph $G_A$};
\end{scope}

\end{tikzpicture}
\caption{Illustration of Example 2. Red dashed edges indicate added edges.}
\label{finger2}
\end{figure}
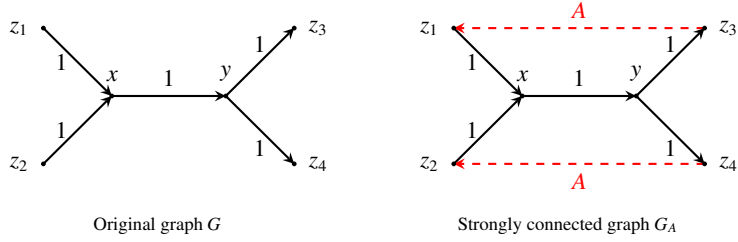

{\it Example 2.} Let $G=(V,E,\mathbf{w})$ be a directed graph, where $V=\{x,y,z_1,z_2,z_3,z_4\}$, $E=\{xy,z_1x,z_2x,yz_3,yz_4\}$, $\mathbf{w}=(1,1,1,1,1)$.
Then $G$ is not strongly connected.
Its strongly connected components are $\{x\}, \{y\}, \{z_1\}, \{z_2\}, \{z_3\}, \{z_4\}$. The condensed graph has sources $\{z_1,z_2\}$ and sinks $\{z_3,z_4\}$. Applying the minimal edge addition procedure described above, we add edge $z_3z_1$ and $z_4z_2$ to obtain a strongly connected graph. For any real number $A>0$, denote the resulting strongly connected graph by $G_A = (V, E_A, \mathbf{w}_A)$, where
$E_A = \{ xy, z_1x,  z_2x,  yz_3, yz_4, z_3z_1, z_4z_2 \}$,
and the edge weights are $w_{A,xy} = w_{A,z_1x} = w_{A,z_2x} = w_{A,yz_3} = w_{A,yz_4} = 1$, $w_{A,z_3z_1} = w_{A,z_4z_2}= A$. Figure~\ref{finger2} shows the original graph $G$ and the strongly connected graph $G_A$.

\subsection{Finding core subgraphs via Ricci curvature flow}
Noting that Ricci curvature flow contracts tightly connected regions while stretching loosely connected ones, we can identify core subgraphs through this process. The following examples illustrate the procedure.\\

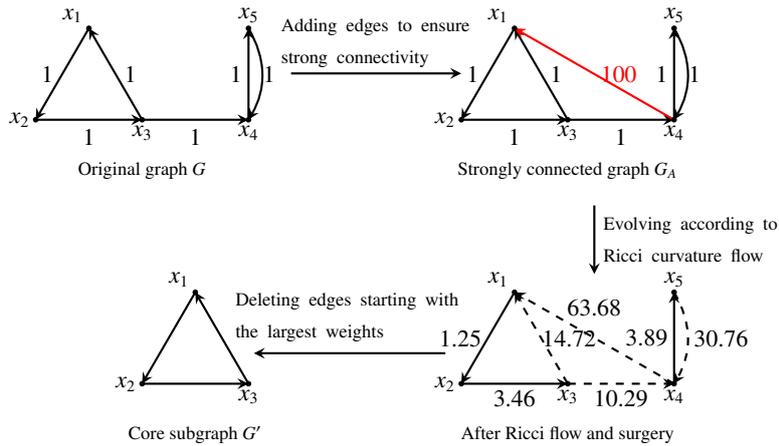
\begin{figure}[H]
\centering
\begin{tikzpicture}[->, >=stealth, scale=0.7, every node/.style={inner sep=0pt, minimum size=0pt}]

\begin{scope}[xshift=0cm, yshift=0cm]
\coordinate (x1) at (0,1.1547);
\coordinate (x2) at (-1,-0.5774);
\coordinate (x3) at (1,-0.5774);
\coordinate (x4) at (3,-0.5774);
\coordinate (x5) at (3,1.1547);

\draw[->, thick] (x1) -- (x2) node[midway,left,xshift=-3pt] {\small 1};
\draw[->, thick] (x2) -- (x3) node[midway,below,yshift=-3pt] {\small 1};
\draw[->, thick] (x3) -- (x1) node[midway,right,xshift=3pt] {\small 1};
\draw[->, thick] (x3) -- (x4) node[midway,below,yshift=-3pt] {\small 1};
\draw[->, thick] (x4) -- (x5) node[midway,left,xshift=-2pt] {\small 1};
\draw[bend left=30, ->, thick] (x5) to node[midway,right] {\small 1} (x4);

\foreach \point in {x1,x2,x3,x4,x5}{ \fill (\point) circle (1.5pt); }

\node[above left=2pt] at (x1) {\small $x_1$};
\node[left=2pt] at (x2) {\small $x_2$};
\node[below=2pt] at (x3) {\small $x_3$};
\node[below=2pt] at (x4) {\small $x_4$};
\node[above=2pt] at (x5) {\small $x_5$};

\node[below=0.4cm] at (1.0,-0.8) {\scriptsize Original graph $G$};
\end{scope}

\draw[->, thick] (3.8,0.3) -- (7,0.3) 
  node[midway, above, yshift=2pt, text width=2.5cm, align=left] 
  {\scriptsize Adding edges to ensure strong connectivity};

\begin{scope}[xshift=8cm, yshift=0cm]
\coordinate (x1) at (0,1.1547);
\coordinate (x2) at (-1,-0.5774);
\coordinate (x3) at (1,-0.5774);
\coordinate (x4) at (3,-0.5774);
\coordinate (x5) at (3,1.1547);

\draw[->, thick] (x1) -- (x2) node[midway,left,xshift=-3pt] {\small 1};
\draw[->, thick] (x2) -- (x3) node[midway,below,yshift=-3pt] {\small 1};
\draw[->, thick] (x3) -- (x1) node[midway,right,xshift=3pt] {\small 1};
\draw[->, thick] (x3) -- (x4) node[midway,below,yshift=-3pt] {\small 1};
\draw[->, thick] (x4) -- (x5) node[midway,left,xshift=-2pt] {\small 1};
\draw[bend left=30, ->, thick] (x5) to node[midway,right] {\small 1} (x4);
\draw[red, thick, ->] (x4) -- node[midway,right,xshift=2pt] {\small 100} (x1);

\foreach \point in {x1,x2,x3,x4,x5}{ \fill (\point) circle (1.5pt); }

\node[above left=2pt] at (x1) {\small $x_1$};
\node[left=2pt] at (x2) {\small $x_2$};
\node[below=2pt] at (x3) {\small $x_3$};
\node[below=2pt] at (x4) {\small $x_4$};
\node[above=2pt] at (x5) {\small $x_5$};

\node[below=0.4cm] at (1.0,-0.8) {\scriptsize Strongly connected graph $G_A$};
\end{scope}

\begin{scope}[xshift=8cm, yshift=-5cm]
\coordinate (x1) at (0,1.1547);
\coordinate (x2) at (-1,-0.5774);
\coordinate (x3) at (1,-0.5774);
\coordinate (x4) at (3,-0.5774);
\coordinate (x5) at (3,1.1547);

\draw[->, thick] (x1) -- (x2) node[midway,left,xshift=-2pt] {\small 1.25};
\draw[->, thick] (x2) -- (x3) node[midway, below,yshift=-2pt] {\small 3.46};
\draw[->, thick] (x4) -- (x5) node[midway,left,xshift=-2pt] {\small 3.89};

\draw[dashed, ->, thick] (x3) -- (x1) node[midway,right] {\small 14.72};
\draw[dashed, ->, thick] (x3) -- (x4) node[midway,below,yshift=-2pt] {\small 10.29};
\draw[dashed, ->, thick] (x4) -- (x1) node[midway,above, yshift=8pt] {\small 63.68};
\draw[dashed,bend left=30, ->, thick] (x5) to node[midway,right,xshift=2pt] {\small 30.76} (x4);

\foreach \point in {x1,x2,x3,x4,x5}{ \fill (\point) circle (1.5pt); }

\node[above left=2pt] at (x1) {\small $x_1$};
\node[left=2pt] at (x2) {\small $x_2$};
\node[below=2pt] at (x3) {\small $x_3$};
\node[below=2pt] at (x4) {\small $x_4$};
\node[above=2pt] at (x5) {\small $x_5$};

\node[below=0.4cm] at (1.0,-0.8) {\scriptsize After Ricci flow and surgery};
\end{scope}

\begin{scope}[xshift=2cm, yshift=-5cm]
\coordinate (x1) at (0,1.1547);
\coordinate (x2) at (-1,-0.5774);
\coordinate (x3) at (1,-0.5774);

\draw[->, thick] (x1) -- (x2);
\draw[->, thick] (x2) -- (x3);
\draw[->, thick] (x3) -- (x1);

\foreach \point in {x1,x2,x3}{ \fill (\point) circle (1.5pt); }

\node[above left=2pt] at (x1) {\small $x_1$};
\node[left=2pt] at (x2) {\small $x_2$};
\node[below=2pt] at (x3) {\small $x_3$};

\node[below=0.4cm] at (0,-0.8) {\scriptsize Core subgraph $G^{\prime}$};
\end{scope}

\draw[->, thick] (9.5,-2.2) -- (9.5,-3.5) node[midway,right, xshift=3pt, text width=2.5cm, align=left] {\scriptsize Evolving according to Ricci curvature flow};

\draw[->, thick] (6.7,-5) -- (3.1,-5) node[midway, above, yshift=4pt, text width=3cm, align=left] {\scriptsize Deleting edges starting with the largest weights};

\end{tikzpicture}

\caption{Illustration of Example 3.}
\label{fig1}
\end{figure}

{\it Example 3.}
Consider the weighted directed graph $G=(V, E, \omega)$ shown in Figure~\ref{fig1}, where $V=\{x_1,x_2,x_3,x_4,x_5\}$, $E=\{x_1x_2, x_2x_3, x_3x_1, x_3x_4, x_4x_5, x_5x_4\}$
with edge weights $\omega_{x_1x_2}=\omega_{x_2x_3}=\omega_{x_3x_1}=\omega_{x_3x_4}=\omega_{x_4x_5}=\omega_{x_5x_4}=1$.
Since $G$ is not strongly connected, we first add edge $x_4x_1$ with a large weight of $100$, obtaining the strongly connected graph $G_A=(V, E_A, \omega)$, where $E_A=E \cup \{x_4x_1\}$, $\omega_{x_4x_1}=100$.
We then apply the discrete Ricci curvature flow (\ref{disc-flow}) to $G_A$ with parameters $n=5$, $\alpha=0.1$, and step size $s=0.1$.
After the flow, the edge deletion strategy removes edges $x_4x_1$, $x_5x_4$, $x_3x_1$ and $x_3x_4$.
The remaining nodes induce a subgraph of the original graph, from which the largest strongly connected component is selected as the core subgraph $G^{\prime}=(V^{\prime}, E^{\prime})$. Here, $V^{\prime}= \{x_1, x_2, x_3\}$ and the edge set $E^{\prime} = \{x_1x_2,  x_2x_3, x_3x_1\}$.
It follows that $r_d^{\mathrm{in}}=1$, $r_d^{\mathrm{out}}=5/6$, and $r_s=1$.

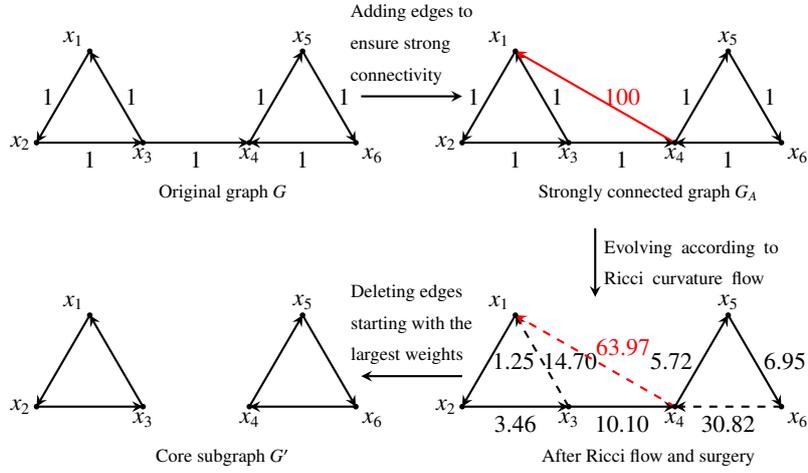
\begin{figure}[H]
\centering
\begin{tikzpicture}[->, >=stealth, scale=0.7, every node/.style={inner sep=0pt, minimum size=0pt}]

\begin{scope}[xshift=0cm, yshift=0cm]
\coordinate (x1) at (0,1.1547);
\coordinate (x2) at (-1,-0.5774);
\coordinate (x3) at (1,-0.5774);
\coordinate (x4) at (3,-0.5774);
\coordinate (x5) at (4,1.1547);
\coordinate (x6) at (5,-0.5774);

\draw[->, thick] (x1) -- (x2) node[midway,left,xshift=-3pt] {\small 1};
\draw[->, thick] (x2) -- (x3) node[midway,below,yshift=-3pt] {\small 1};
\draw[->, thick] (x3) -- (x1) node[midway,right,xshift=3pt] {\small 1};

\draw[->, thick] (x4) -- (x5) node[midway,left,xshift=-3pt] {\small 1};
\draw[->, thick] (x5) -- (x6) node[midway,right,xshift=3pt] {\small 1};
\draw[->, thick] (x6) -- (x4) node[midway,below,yshift=-3pt] {\small 1};

\draw[->, thick] (x3) -- (x4) node[midway,below,yshift=-3pt] {\small 1};

\foreach \point in {x1,x2,x3,x4,x5,x6}{ \fill (\point) circle (1.5pt); }

\node[above left=2pt] at (x1) {\small $x_1$};
\node[left=2pt] at (x2) {\small $x_2$};
\node[below=2pt] at (x3) {\small $x_3$};
\node[below=2pt] at (x4) {\small $x_4$};
\node[above=2pt] at (x5) {\small $x_5$};
\node[below right=2pt] at (x6) {\small $x_6$};

\node[below=0.4cm] at (2.5,-0.8) {\scriptsize Original graph $G$};
\end{scope}

\draw[->, thick] (5.1,0.3) -- (7,0.3) 
  node[midway, above, yshift=4pt, align=left] 
  {\scriptsize Adding edges to\\
   \scriptsize ensure strong\\
   \scriptsize connectivity};

\begin{scope}[xshift=8cm, yshift=0cm]
\coordinate (x1) at (0,1.1547);
\coordinate (x2) at (-1,-0.5774);
\coordinate (x3) at (1,-0.5774);
\coordinate (x4) at (3,-0.5774);
\coordinate (x5) at (4,1.1547);
\coordinate (x6) at (5,-0.5774);

\draw[->, thick] (x1) -- (x2) node[midway,left,xshift=-3pt] {\small 1};
\draw[->, thick] (x2) -- (x3) node[midway,below,yshift=-3pt] {\small 1};
\draw[->, thick] (x3) -- (x1) node[midway,right,xshift=3pt] {\small 1};

\draw[->, thick] (x4) -- (x5) node[midway,left,xshift=-3pt] {\small 1};
\draw[->, thick] (x5) -- (x6) node[midway,right,xshift=3pt] {\small 1};
\draw[->, thick] (x6) -- (x4) node[midway,below,yshift=-3pt] {\small 1};

\draw[->, thick] (x3) -- (x4) node[midway,below,yshift=-3pt] {\small 1};

\draw[red, thick, ->] (x4) -- node[midway,right,xshift=3pt] {\small 100} (x1);

\foreach \point in {x1,x2,x3,x4,x5,x6}{ \fill (\point) circle (1.5pt); }

\node[above left=2pt] at (x1) {\small $x_1$};
\node[left=2pt] at (x2) {\small $x_2$};
\node[below=2pt] at (x3) {\small $x_3$};
\node[below=2pt] at (x4) {\small $x_4$};
\node[above=2pt] at (x5) {\small $x_5$};
\node[below right=2pt] at (x6) {\small $x_6$};

\node[below=0.4cm] at (2.5,-0.8) {\scriptsize Strongly connected graph $G_A$};
\end{scope}

\begin{scope}[xshift=0cm, yshift=-5cm]
\coordinate (x1) at (0,1.1547);
\coordinate (x2) at (-1,-0.5774);
\coordinate (x3) at (1,-0.5774);

\coordinate (x4) at (3,-0.5774);
\coordinate (x5) at (4,1.1547);
\coordinate (x6) at (5,-0.5774);

\draw[->, thick] (x1) -- (x2);
\draw[->, thick] (x2) -- (x3);
\draw[->, thick] (x3) -- (x1);

\draw[->, thick] (x4) -- (x5);
\draw[->, thick] (x5) -- (x6);
\draw[->, thick] (x6) -- (x4);

\foreach \point in {x1,x2,x3,x4,x5,x6}{ \fill (\point) circle (1.5pt); }

\node[above left=2pt] at (x1) {\small $x_1$};
\node[left=2pt] at (x2) {\small $x_2$};
\node[below=2pt] at (x3) {\small $x_3$};
\node[below=2pt] at (x4) {\small $x_4$};
\node[above=2pt] at (x5) {\small $x_5$};
\node[below right=2pt] at (x6) {\small $x_6$};

\node[below=0.4cm] at (2.5,-0.8) {\scriptsize Core subgraph $G^{\prime}$};
\end{scope}

\begin{scope}[xshift=8cm, yshift=-5cm]
\coordinate (x1) at (0,1.1547);
\coordinate (x2) at (-1,-0.5774);
\coordinate (x3) at (1,-0.5774);
\coordinate (x4) at (3,-0.5774);
\coordinate (x5) at (4,1.1547);
\coordinate (x6) at (5,-0.5774);

\draw[->, thick] (x1) -- (x2) node[midway, right,xshift=1pt] {\small 1.25};
\draw[->, thick] (x2) -- (x3) node[midway,below,yshift=-3pt] {\small 3.46};
\draw[dashed, ->, thick] (x3) -- (x1) node[midway,right] {\small 14.70};

\draw[->, thick] (x4) -- (x5) node[midway,left,xshift=-3pt] {\small 5.72};
\draw[->, thick] (x5) -- (x6) node[midway,right,xshift=3pt] {\small 6.95};
\draw[dashed, ->, thick] (x6) -- (x4) node[midway,below,yshift=-3pt] {\small 30.82};

\draw[->, thick] (x3) -- (x4) node[midway,below,yshift=-3pt] {\small 10.10};

\draw[red, thick, dashed, ->] (x4) -- (x1) node[midway,right,yshift=5pt] {\small 63.97};

\foreach \point in {x1,x2,x3,x4,x5,x6}{ \fill (\point) circle (1.5pt); }

\node[above left=2pt] at (x1) {\small $x_1$};
\node[left=2pt] at (x2) {\small $x_2$};
\node[below=2pt] at (x3) {\small $x_3$};
\node[below=2pt] at (x4) {\small $x_4$};
\node[above=2pt] at (x5) {\small $x_5$};
\node[below right=2pt] at (x6) {\small $x_6$};

\node[below=0.4cm] at (2.5,-0.8) {\scriptsize After Ricci flow and surgery};
\end{scope}

\draw[->, thick] (9.5,-2.2) -- (9.5,-3.5) node[midway,right, xshift=3pt, text width=2.5cm, align=left] {\scriptsize Evolving according to Ricci curvature flow};

\draw[->, thick] (7,-5) -- (5.1,-5) node[midway, above,yshift=4pt, align=left] {\scriptsize Deleting edges\\ \scriptsize starting with the\\
\scriptsize largest weights};

\end{tikzpicture}

\caption{Illustration of Example 4.}
\label{fig2}
\end{figure}

{\it Example 4.}
Consider the weighted directed graph $G=(V,E,\omega)$ shown in Figure~\ref{fig2}, 
where $V=\{x_1,x_2,x_3,x_4,x_5,x_6\}$ and $E=\{x_1x_2,\,x_2x_3,\,x_3x_1,\,x_3x_4,\,x_4x_5,\,x_5x_6,\,x_6x_4\}$, and each edge initially has weight $1$. 
Since $G$ is not strongly connected, we first add the edge $x_4x_1$ with a large weight of $100$, 
obtaining the strongly connected graph $G_A=(V,E_A,\omega)$ with 
$E_A = E \cup \{x_4x_1\}$ and $\omega_{x_4x_1}=100$. 
We then apply the discrete Ricci curvature flow (\ref{disc-flow}) to $G_A$ with parameters $n=5$, $\alpha=0.1$, and step size $s=0.1$. 
After the flow, the edge deletion strategy removes the edges $x_4x_1$, $x_6x_4$, and $x_3x_1$. 
The remaining nodes induce a subgraph of the original graph, from which the largest strongly connected components are selected as the core subgraph $G'=(V',E')$. 
The final core subgraph consists of two directed triangles,
where $V^{\prime}=\{x_1,x_2,x_3,x_4,x_5,x_6\}$ and $E^{\prime}=\{x_1x_2, x_2x_3, x_3x_1, x_4x_5, x_5x_6, x_6x_4\}$.
It follows that $r_d^{\mathrm{in}}=11/12$, $r_d^{\mathrm{out}}=11/12$, while $r_s$ is not defined.

\subsection{Algorithms}
The algorithm for finding core subgraphs of directed graphs via Ricci curvature flow is as follows.
First, the original graph is preprocessed to ensure strong connectivity by adding artificial edges with large weights. Next, edge weights are iteratively updated according to the discrete Ricci curvature flow equation (\ref{disc-flow}) with a fixed step size $s$ and curvature parameter $\alpha$, over the number of iterations $N$. After the flow, edges are sorted by their updated weights in descending order, and the top $\tau\%$ of edges are removed. Any isolated nodes resulting from this deletion are also removed. The remaining nodes induce a subgraph from original graph.
Finally, the largest strongly connected component of this induced subgraph is selected as the detected core subgraph.
The corresponding pseudo-code is provided in Algorithm~\ref{algorithm-directed}.\\

\begin{algorithm}[H]
\caption{Finding core subgraphs via Ricci curvature flow on directed graphs}
\label{algorithm-directed}
\KwIn{Directed weighted graph $G=(V,E,w)$; maximum iteration $N$; edge removal ratio $\tau$; step size $s$; curvature parameter $\alpha$.}
\KwOut{Core subgraph $G^{\prime}$.}

\textbf{Step 1: Add edges to ensure strong connectivity}\;
Decompose $G$ into strongly connected components\;
Add artificial edges with large weight $A>0$ to connect the components in a cycle\;
Mark all added edges as \textit{artificial}\;
Ensure that the resulting graph $G_A$ is strongly connected\;

\textbf{Step 2: Evolve according to Ricci curvature flow}\;
\For{$i \gets 0$ \KwTo $N-1$}{
    Update $w_e$ for all $e\in E$ using Ricci curvature flow (\ref{disc-flow}) with step size $s$ and curvature parameter $\alpha$\;
}

\textbf{Step 3: Remove artificial edges and delete edges starting with the largest weights}\;
Delete all artificial edges that were added in Step 1\;
Sort remaining edges by $w_e^{(N)}$ in ascending order and retain the top $(1-\tau)\%$ of edges\;
Remove all nodes that become isolated after the edge removal process\;

\textbf{Step 4: Construct candidate core subgraph}\;
Let $\mathcal{S}$ be the set of remaining nodes\;
Construct the induced subgraph $G_{\mathcal{S}} = G[\mathcal{S}]$ from $G$\;

\textbf{Step 5: Find core subgraph}\;
Identify the largest strongly connected component of $G_{\mathcal{S}}$
and denote it by $G^{\prime}$\;

Return $G^{\prime}$ as the detected core subgraph\;
\end{algorithm}

\vspace{1em} 

\noindent\textbf{Remark.}
A degenerate case may occur when the induced subgraph $G_{\mathcal{S}}$ contains no nontrivial strongly connected components. In this case, each node forms a trivial component by itself, and the algorithm will output isolated nodes as the core subgraph.
This is a theoretical possibility inherent to directed graphs. Nevertheless, this situation rarely occurs in real-world directed networks.

\vspace{1em} 

The time complexity of the proposed core detection algorithm on directed graphs is dominated by the Ricci curvature flow phase. In each of the $N$ iterations, edge weights are updated according to their discrete Ricci curvature. Computing Ricci curvature for a single edge involves solving an optimal transport problem, which incurs a cost of $\mathcal{O}(D^3)$, where $D$ denotes the average degree of nodes. Consequently, each iteration takes $\mathcal{O}(|E| D^3)$ time, resulting in an overall complexity of $\mathcal{O}(N |E| D^3)$ for the Ricci curvature flow phase.
The subsequent steps include sorting the final edge weights in descending order, which requires $\mathcal{O}(|E| \log |E|)$, removing isolated nodes and constructing the induced subgraph. Identifying the largest strongly connected component takes $\mathcal{O}(|V| + |E|)$ time.
Overall, the algorithm has a total time complexity of $\mathcal{O}(N |E| D^3 + |E| \log |E| + |V| + |E|)$, where in practice the cubic dependence on the average degree $D$ makes the term $\mathcal{O}(N |E| D^3)$ the dominant contributor to the computational cost.

\section{Experiments}
In this section, we present experiments on three real-world directed networks to evaluate the performance of the Ricci curvature flow method. 
We first extract core subgraphs using Algorithm~\ref{algorithm-directed} with specific parameters, and then assess their structural properties. 
Finally, we compare the extracted cores with those obtained from several baseline centrality measures to demonstrate the effectiveness of our approach.

\subsection{Real-world Datasets}
Basic information for three real-world directed networks are listed in Table~\ref{data}.

\begin{table}[H]
\centering
\caption{Statistical properties of the networks used in our analysis. If the original network is disconnected, we only consider its largest weak connected component.}
\label{data}
\begin{tabular}{lccccc}
\toprule
Network & Vertices & Edges & AvgDeg & Diameter & Density  \\
\midrule
Physicians & 117 & 542 & 9.26 & 5 & 0.040 \\
Elegans   & 297 & 2345 & 15.79 & 5 & 0.027  \\
Human protein & 1615 & 6105 & 7.56 & 13 & 0.002 \\
\bottomrule
\end{tabular}
\end{table}

The Physicians network \cite{human} represents professional interactions among 246 physicians across four towns in Illinois: Peoria, Bloomington, Quincy, and Galesburg. The network captures communication and influence relationships between physicians. Its largest weakly connected component consists of 117 nodes and 542 edges.
The Elegans network \cite{elegans} captures functional associations among 297 genes in Caenorhabditis elegans.
Each node represents a gene, edges represent predicted functional relationships based on multiple biological data sources. The network includes 2345 edges.
The Human protein network \cite{human} contains 1615 nodes and 6105 edges, where each node represents a human protein and each edge indicates a human binary protein-protein interactions.

For all experiments, we set the step size of the Ricci curvature flow to $s = 0.1$, consistent with the valid range $0 < s < 1$ established by Theorem \ref{discrete-thm}. The edge removal ratio is fixed at $\tau = 80\%$, meaning that the top 80\% of edges are removed after the final iteration of the Ricci curvature flow. The number of iterations $N$ and the curvature parameter $\alpha$ are selected for each dataset to optimize structural metrics, with $\alpha$ further investigated in the next subsection to examine its specific effects.

\subsection{Effect of curvature parameter $\alpha$}
\begin{figure}[H]
    \centering
    \includegraphics[width=0.46\textwidth]{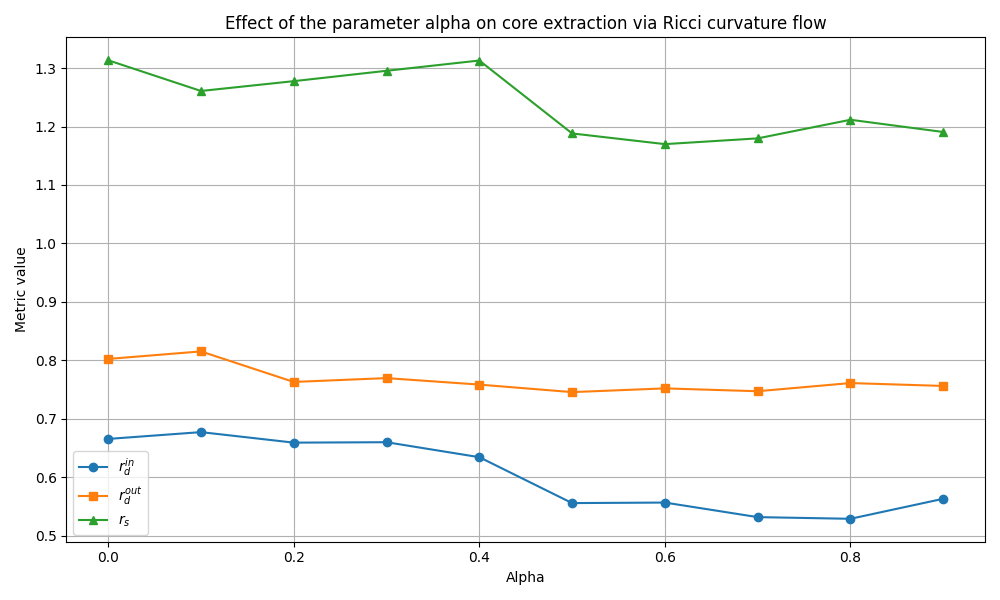}
    \caption{Effect of curvature parameter $\alpha$ on the Physicians network.}
    \label{fig:alpha_physicians}
\end{figure}

\begin{figure}[H]
    \centering
    \includegraphics[width=0.46\textwidth]{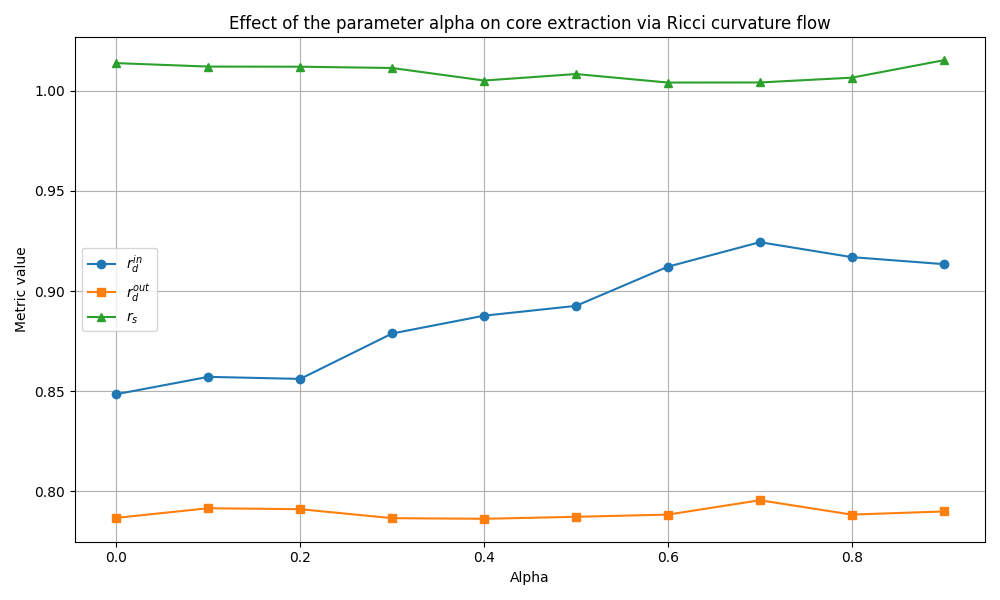}
    \caption{Effect of curvature parameter $\alpha$ on the Elegans network.}
    \label{fig:alpha_elegans}
\end{figure}

\begin{figure}[H]
    \centering
    \includegraphics[width=0.46\textwidth]{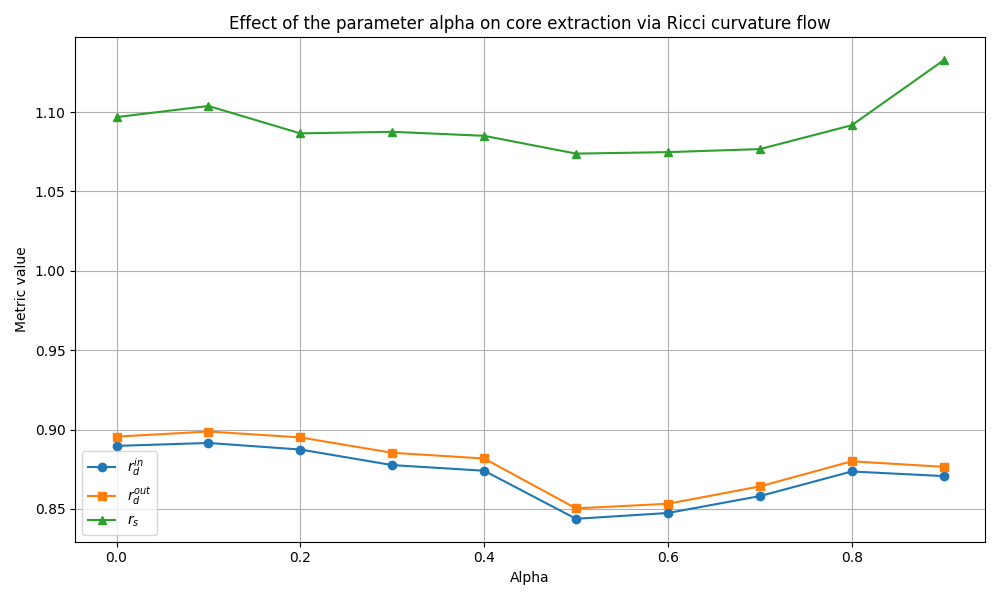}
    \caption{Effect of curvature parameter $\alpha$ on the Human protein network.}
    \label{fig:alpha_humanprotein}
\end{figure}

Note that the performance of the Ricci curvature flow method depends on the choice of the curvature parameter $\alpha$,  which in turn affects the selection of core subgraph. 
To systematically investigate its impact, we vary $\alpha$ in the range $[0,1)$ and evaluate three structural metrics: in-degree core cohesion $r_d^{\mathrm{in}}$, out-degree core cohesion $r_d^{\mathrm{out}}$, and average distance stretch $r_s$. 
For these experiments, the number of iterations $N$ has been set to optimize the structural metrics for each dataset: $N=30$ for Physicians, $N=5$ for Elegans, and $N=50$ for Human protein.
The experimental results are presented in Figures~\ref{fig:alpha_physicians}-\ref{fig:alpha_humanprotein}, which illustrate how different values of $\alpha$ influence the cohesiveness and structural centrality of the extracted core subgraphs.

For the Physicians network, both $r_d^{\mathrm{in}}$ and $r_d^{\mathrm{out}}$ reach their highest values at $\alpha = 0.1$ and gradually decline as $\alpha$ increases, indicating a reduced core cohesion for larger $\alpha$. In contrast, $r_s$ reaches its maximum at $\alpha = 0$, remains slightly lower at $\alpha = 0.1$, and then generally decreases with minor fluctuations, showing a small rebound near $\alpha = 0.9$. Overall, these patterns suggest that a small but nonzero $\alpha$ enables the Ricci curvature flow to balance local connectivity with global structural influence, resulting in cores that are both cohesive and topologically central.

For the Elegans network, both $r_d^{\mathrm{in}}$ and $r_d^{\mathrm{out}}$ increase slightly from $\alpha = 0$ to $\alpha = 0.7$, reaching their highest values around $\alpha = 0.7$, and then show a mild decline as $\alpha$ approaches $0.9$. The structural stretch ratio $r_s$ remains close to $1$ across all $\alpha$, with a slight increase at $\alpha = 0.9$. This suggests that moderate values of $\alpha$ enable the Ricci curvature flow to balance local and global effects, generating cores that are both compact and topologically influential.

For the Human protein network, both $r_d^{\mathrm{in}}$ and $r_d^{\mathrm{out}}$ remain relatively stable for small $\alpha$, reaching their peak at $\alpha = 0.1$, and gradually decline afterward, indicating that higher $\alpha$ values reduce local structural cohesion. The value of $r_s$ also attains a local maximum at $\alpha = 0.1$, slightly higher than at $\alpha = 0$, before fluctuating and rising again near $\alpha = 0.9$. These results suggest that $\alpha = 0.1$ provides an optimal balance between local and global structural effects in the Ricci curvature flow, yielding core subgraphs that are both cohesive and structurally representative.

\subsection{Core extraction via Ricci curvature flow}
Using Algorithm~\ref{algorithm-directed}, we extract the core subgraphs from each network based on the selected parameters. The number of iterations $N$ and the curvature parameter $\alpha$ are set for each dataset to optimize the three structural metrics: $N = 30$, $\alpha=0.1$ for the Physicians network, $N = 5$, $\alpha = 0.9$ for the Elegans network, and $N = 50$, $\alpha = 0.1$ for the Human protein network. Table~\ref{tab:ricci_core} summarizes the number of core nodes and edges for each network, along with the corresponding structural metrics.

\begin{table}[H]
\centering
\caption{Core extraction results of Ricci curvature flow across three networks}
\label{tab:ricci_core}
\begin{tabular}{lccccccc}
\toprule
Network & \#Original nodes & \#Core nodes & \#Core edges & $r_d^{\mathrm{in}}$ & $r_d^{\mathrm{out}}$ & $r_s$ \\
\midrule
Physicians       & 117  & 59  & 242  & 0.6771 & 0.8152 & 1.2609 \\
Elegans          & 297  & 193 & 1583 & 0.9135 & 0.7900 & 1.0153 \\
Human protein    & 1615 & 537 & 1898 & 0.8853 & 0.8925 & 1.0913 \\
\bottomrule
\end{tabular}
\end{table}

The results in Table~\ref{tab:ricci_core} show that the Ricci curvature flow method effectively identifies cohesive and structurally significant core subgraphs across all three networks.
For the Physicians network, the extracted core consists of 59 nodes and 242 edges, forming a compact and highly interactive cluster that preserves both internal and external connectivity, with in-degree core cohesion 0.6771 and out-degree core cohesion 0.8152. The average distance stretch is 1.2609, indicating that the extracted core maintains efficient communication while concentrating structural importance.
For the Elegans network, the flow converges rapidly within five iterations and reveals a biologically meaningful core of 193 nodes and 1583 edges, maintaining a balance between local density and global reach. The corresponding structural metrics are 0.9135 for in-degree core cohesion, 0.7900 for out-degree core cohesion, and 1.0153 for average distance stretch.
For the Human protein network, the extracted core contains 537 proteins and 1898 interactions, representing a densely interconnected backbone that preserves essential signaling pathways. The in-degree and out-degree core cohesion ratios are 0.8853 and 0.8925, respectively, and the average distance stretch is 1.0913.
Overall, these results demonstrate that the Ricci curvature flow dynamically concentrates structural importance into a smaller, well-connected subgraph. The identified cores not only retain strong internal cohesion but also maintain the overall structural  efficiency of the original networks.

\begin{figure}[H]
    \centering
    \begin{subfigure}[b]{0.32\textwidth}
        \centering
        \includegraphics[width=\textwidth]{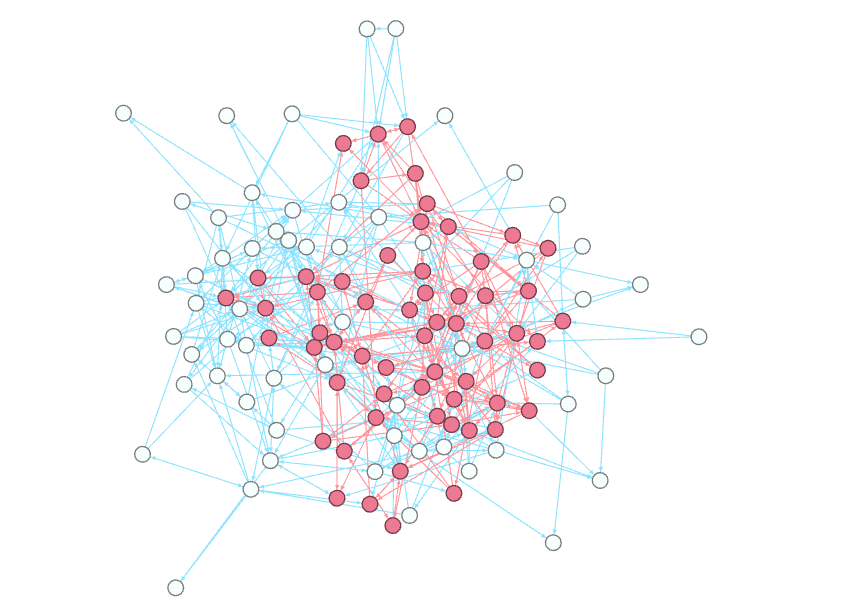}
        \caption{Physicians network}
        \label{fig:core_physicians}
    \end{subfigure}
    \hfill
    \begin{subfigure}[b]{0.32\textwidth}
        \centering
        \includegraphics[width=\textwidth]{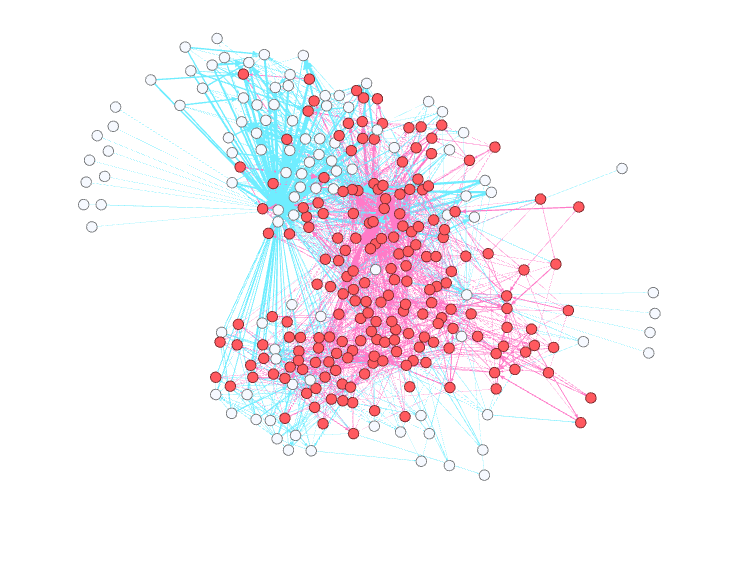}
        \caption{Elegans network}
        \label{fig:core_elegans}
    \end{subfigure}
    \hfill
    \begin{subfigure}[b]{0.32\textwidth}
        \centering
        \includegraphics[width=\textwidth]{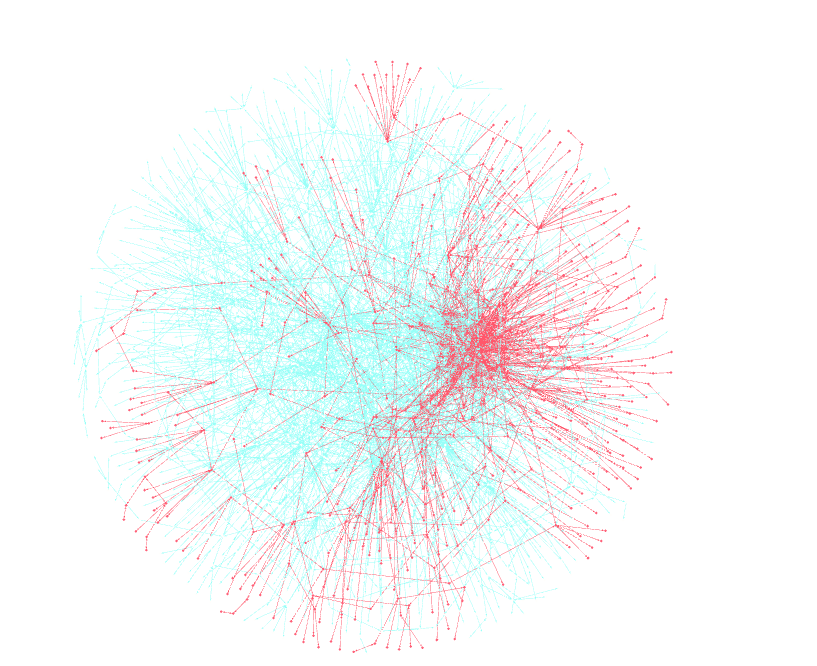}
        \caption{Human protein network}
        \label{fig:core_humanprotein}
    \end{subfigure}
    \caption{Visualization of the core subgraphs extracted by the Ricci curvature flow method across three networks.}
    \label{fig:core_all}
\end{figure}

Figure~\ref{fig:core_all} shows the core subgraphs extracted by the Ricci curvature flow method for the Physicians, Elegans, and Human protein networks. Core nodes are highlighted in red and non-core nodes in white. Edges between core nodes can be emphasized in Gephi to illustrate the internal structure of each core.

\subsection{Comparison with baseline centrality methods}
To evaluate the performance of Algorithm~\ref{algorithm-directed} on directed graphs, we compare it against four commonly used node centrality measures: degree, betweenness, closeness, and Pagerank.
For degree centrality, nodes are ranked based on the sum of their in-degree and out-degree, highlighting nodes with the most connections.
Betweenness centrality identifies nodes that frequently appear on shortest paths between other node pairs,
emphasizing nodes that act as bridges in the network.
Closeness centrality is computed in the outward direction,
reflecting how efficiently a node can reach other nodes.
Pagerank assigns scores to nodes according to the stationary distribution of a random walk, with higher scores for nodes connected to other highly ranked nodes. For precise definitions and a more detailed discussion of these centrality measures, readers are referred to \cite{freeman, newman2, page}.
For each centrality measure, the top-ranked nodes are selected as candidate core nodes, and the subgraph induced by these nodes is extracted. The largest strongly connected component of this subgraph is then identified as the final core subgraph. The number of nodes selected in each baseline method is set to match the size of the core subgraph extracted by the Ricci curvature flow algorithm to ensure a fair comparison.

To quantify the structural properties of the detected core subgraphs, we compute three metrics: the in-degree core cohesion $r_d^{\mathrm{in}}$,
the out-degree core cohesion $r_d^{\mathrm{out}}$,
and the average distance stretch $r_s$ after removing the core subgraph. The comparison results of core subgraph extraction methods on the Physicians, Elegans and Human protein datasets are presented in Tables \ref{tab2}-\ref{tab4}, respectively.

\begin{table}[H]
\centering
\caption{Comparison of core extraction methods on the Physicians network}
\label{tab2}
\begin{tabular}{lccccc}
\toprule
Method & \#Core nodes & \#Core edges & $r_d^{\mathrm{in}}$ & $r_d^{\mathrm{out}}$ & $r_s$ \\
\midrule
Ricci flow              & 59 & 242 & \textbf{0.6771} & 0.8152 & \textbf{1.2609}\\
Pagerank                & 59 & 253 & 0.6119 & 0.8544 & 1.1988 \\
Degree centrality       & 59 & 265 & 0.6312 & 0.7956 & 1.0175 \\
Betweenness centrality  & 59 & 244 & 0.5854 & 0.7596 & 1.0000 \\
Closeness centrality    & 59 & 257 & 0.6303 & \textbf{0.8721} & 1.1433 \\
\bottomrule
\end{tabular}
\begin{flushleft}
\end{flushleft}
\end{table}

\begin{table}[H]
\centering
\caption{Comparison of core extraction methods on the Elegans network}
\label{tab3}
\begin{tabular}{lccccc}
\toprule
Method & \#Core nodes & \#Core edges & $r_d^{\mathrm{in}}$ & $r_d^{\mathrm{out}}$ & $r_s$ \\
\midrule
Ricci flow              & 193 & 1583 & \textbf{0.9135} & \textbf{0.7900} & \textbf{1.0153}\\
Pagerank                & 193 & 1546 & 0.8268 & 0.7705 & 1.0057 \\
Degree centrality       & 193 & 1668 & 0.8993 & 0.7734 & 1.0100 \\
Betweenness centrality  & 193 & 1590 & 0.8721 & 0.7754 & 1.0069 \\
Closeness centrality    & 193 & 1553 & 0.8372 & 0.7484 & 1.0090 \\
\bottomrule
\end{tabular}
\begin{flushleft}
\end{flushleft}
\end{table}

\begin{table}[H]
\centering
\caption{Comparison of core extraction methods on the Human protein network}
\label{tab4}
\begin{tabular}{lccccc}
\toprule
Method & \#Core nodes & \#Core edges & $r_d^{\mathrm{in}}$ & $r_d^{\mathrm{out}}$ & $r_s$ \\
\midrule
Ricci flow              & 537 & 1898 & \textbf{0.8853} & \textbf{0.8925} & 1.0913\\
Pagerank                & 537 & 3336 & 0.7437 & 0.7368 & 1.0000 \\
Degree centrality       & 537 & 3542 & 0.7776 & 0.7681 & 1.0000 \\
Betweenness centrality  & 537 & 3107 & 0.7168 & 0.7081 & 1.0000 \\
Closeness centrality    & 537 & 3405 & 0.8252 & 0.8198 & \textbf{1.4401} \\
\bottomrule
\end{tabular}
\begin{flushleft}
\end{flushleft}
\end{table}

The results presented in Tables~\ref{tab2}-\ref{tab4} show that our method consistently identifies core subgraphs with superior structural cohesiveness compared to the four baseline centrality measures across all three directed networks. Specifically, on the Physicians dataset, Ricci curvature flow achieves the highest in-degree cohesiveness $r_d^{\mathrm{in}}$ of 0.6771 and the largest average distance stretch $r_s$ of 1.2609, indicating that the extracted core subgraph is both tightly connected and structurally influential.  Similarly, for the Elegans dataset, the Ricci curvature flow method attains the highest $r_d^{\mathrm{in}}$ of 0.9135 and $r_d^{\mathrm{out}}$ of 0.7900, together with the largest $r_s$ of 1.0153, outperforming all four centrality-based baselines. On the Human protein dataset, Ricci curvature flow achieves leading values of $r_d^{\mathrm{in}}$ of 0.8853 and $r_d^{\mathrm{out}}$ of 0.8925.
These results collectively demonstrate that the Ricci curvature flow algorithm effectively extracts core subgraphs that are more cohesive and structurally significant than those obtained using traditional centrality measures.

\subsection{Robustness analysis under core edge deletion}
To comprehensively evaluate the anti-interference capability of core subgraphs extracted by different methods, we first apply Algorithm~\ref{algorithm-directed} to obtain the core of the Human protein network, and then randomly delete different proportions of its core edges, with deletion ratios increasing from $10\%$ to $90\%$.
This procedure allows us to examine how robust the identified cores remain in preserving structural integrity and functional connectivity under progressive edge perturbations.

\begin{table}[H]
  \centering
  \caption{Performance comparison of methods on the Human protein network (core edge deletion ratio:$10\%$ to $30\%$)}
  \label{tab:comparison_10_30}
  \resizebox{\textwidth}{!}{ 
  \begin{tabular}{@{} l | c c c | c c c | c c c @{}}
    \toprule
    & \multicolumn{3}{c|}{$r_d^{\mathrm{in}}$} & \multicolumn{3}{c|}{$r_d^{\mathrm{out}}$} & \multicolumn{3}{c}{$r_s$} \\
    \cmidrule(lr){2-4} \cmidrule(lr){5-7} \cmidrule(lr){8-10}
    \textbf{Method} & $10\%$ & $20\%$ & $30\%$ & $10\%$ & $20\%$ & $30\%$ & $10\%$ & $20\%$ & $30\%$ \\
    \midrule
    Ricci flow & \textbf{0.8289} & \textbf{0.8257} & \textbf{0.8126} & \textbf{0.8362} & \textbf{0.8325} & \textbf{0.8144} & 1.0993 & 1.1499 & 1.1266 \\
    Pagerank   & 0.6928 & 0.6594 & 0.6213 & 0.6770 & 0.6360 & 0.5951 & 1.0004 & 1.0032 & 1.0084 \\
    Degree     & 0.7140 & 0.6505 & 0.5982 & 0.7041 & 0.6432 & 0.5894 & 1.0015 & 1.0000 & 1.0014 \\
    Betweenness& 0.6686 & 0.6323 & 0.6091 & 0.6605 & 0.6280 & 0.6031 & 1.0000 & 1.0003 & 1.0014 \\
    Closeness  & 0.7630 & 0.7156 & 0.6679 & 0.7461 & 0.6926 & 0.6496 & \textbf{1.5198} & \textbf{1.5549} & \textbf{1.8122} \\
    \bottomrule
  \end{tabular}
  }
\end{table}

\begin{table}[H]
  \centering
  \caption{Performance comparison of methods on the Human protein network (core edge deletion ratio:$40\%$ to $60\%$)}
  \label{tab:comparison_40_60}
  \resizebox{\textwidth}{!}{ 
  \begin{tabular}{@{} l | c c c | c c c | c c c @{}}
    \toprule
    & \multicolumn{3}{c|}{$r_d^{\mathrm{in}}$} & \multicolumn{3}{c|}{$r_d^{\mathrm{out}}$} & \multicolumn{3}{c}{$r_s$} \\
    \cmidrule(lr){2-4} \cmidrule(lr){5-7} \cmidrule(lr){8-10}
    \textbf{Method} & $40\%$ & $50\%$ & $60\%$ & $40\%$ & $50\%$ & $60\%$ & $40\%$ & $50\%$ & $60\%$ \\
    \midrule
    Ricci flow & \textbf{0.8192} & \textbf{0.8191} & \textbf{0.8200} & \textbf{0.8204} & \textbf{0.8197} & \textbf{0.8210} & 1.1189 & 1.1091 & 1.0935 \\
    Pagerank   & 0.6013 & 0.5726 & 0.5621 & 0.5739 & 0.5485 & 0.5402 & 1.0308 & 1.2409 & 1.4306 \\
    Degree     & 0.5583 & 0.5378 & 0.5174 & 0.5485 & 0.5370 & 0.5167 & 1.0104 & 1.3374 & 1.4294 \\
    Betweenness& 0.6079 & 0.6099 & 0.6165 & 0.6038 & 0.6026 & 0.6121 & 1.0032 & 1.0036 & 1.0138 \\
    Closeness  & 0.6260 & 0.5956 & 0.5825 & 0.6080 & 0.5765 & 0.5691 & \textbf{1.6522} & \textbf{1.8446} & \textbf{1.9951} \\
    \bottomrule
  \end{tabular}
  }
\end{table}

\begin{table}[H]
  \centering
  \caption{Performance comparison of methods on the Human protein network (core edge deletion ratio:$70\%$ to $90\%$)}
  \label{tab:comparison_70_90}
  \resizebox{\textwidth}{!}{ 
  \begin{tabular}{@{} l | c c c | c c c | c c c @{}}
    \toprule
    & \multicolumn{3}{c|}{$r_d^{\mathrm{in}}$} & \multicolumn{3}{c|}{$r_d^{\mathrm{out}}$} & \multicolumn{3}{c}{$r_s$} \\
    \cmidrule(lr){2-4} \cmidrule(lr){5-7} \cmidrule(lr){8-10}
    \textbf{Method} & $70\%$ & $80\%$ & $90\%$ & $70\%$ & $80\%$ & $90\%$ & $70\%$ & $80\%$ & $90\%$ \\
    \midrule
    Ricci flow & \textbf{0.8216} & \textbf{0.8151} & \textbf{0.8031} & \textbf{0.8226} & \textbf{0.8140} & \textbf{0.8015} & 1.1291 & 1.0917 & 1.1515 \\
    Pagerank   & 0.5546 & 0.5447 & 0.5573 & 0.5371 & 0.5333 & 0.5547 & 1.6702 & 1.6130 & 1.4280 \\
    Degree     & 0.5263 & 0.5566 & 0.6182 & 0.5258 & 0.5545 & 0.6139 & 1.6300 & \textbf{1.8600} & \textbf{1.9095} \\
    Betweenness& 0.6321 & 0.6535 & 0.6870 & 0.6268 & 0.6509 & 0.6859 & 1.0878 & 1.1724 & 1.6665 \\
    Closeness  & 0.5965 & 0.6472 & 0.7606 & 0.5783 & 0.6346 & 0.7546 & \textbf{1.7202} & 1.8132 & 1.1583 \\
    \bottomrule
  \end{tabular}
  }
\end{table}

The experimental results detailed in Tables~\ref{tab:comparison_10_30}-\ref{tab:comparison_70_90} test the robustness of each method by evaluating the extracted core subgraphs under increasing core edge deletion ratios, spanning from 10\% to 90\% on the Human protein network. A systematic analysis of the experimental results across various deletion ratios clearly reveals that the proposed Ricci curvature flow method consistently exhibits superior and highly stable performance on the two core metrics measuring subgraph internal cohesiveness: $r_d^{\mathrm{in}}$ and $r_d^{\mathrm{out}}$. This performance is significantly better than that of other classic methods, including Pagerank, degree, betweenness, and closeness. This consistent advantage demonstrates that the core subgraphs identified by Ricci curvature flow possess the highest intrinsic density and structural stability. Even when the network structure suffers localized damage ranging from mild to extreme, the core cohesiveness of the subgraph remains effectively preserved. For the primary objective of extracting the most internally dense and stable core community, the Ricci curvature flow method exhibits superior and more resilient performance.

We further analyze the network stretch ratio metric $ r_s$. The stretch ratio $r_s$ quantifies the degree of change in the shortest path lengths between nodes within the subgraph after core edge deletion. In the context of network robustness analysis, a crucial interpretation of a low $r_s$ value is its association with minimized communication cost and path resilience: when facing large-scale edge deletion or component failures, a small $r_s$ implies that the change in communication path length between nodes is minimal, meaning the operational or communication cost remains stable. This stability is vital for scenarios demanding high resistance to extreme interference, such as emergency communication networks and critical infrastructure control systems, ensuring that communication efficiency does not sharply decline during a crisis.

It is particularly noteworthy that the distinguishing characteristic of the Ricci curvature flow method on the $r_s$ metric is its steadiness. While other methods, notably closeness and degree, achieve higher $r_s$ values at different deletion ratios, this indicates their relative strength in identifying nodes critical for global network paths. The $r_s$ value of the Ricci curvature flow method fluctuates minimally across the entire $10\%$ to $90\%$ deletion range, consistently clustering around a relatively constant value. This high degree of steadiness signifies that the core structure extracted by Ricci curvature flow has an inherent resistance to continuous network topological degradation, consistently guaranteeing stable communication paths between nodes. It effectively prevents drastic fluctuations in communication costs during extreme interference, thereby ensuring the reliability and continuity of core functionality.

\section{Conclusion}
In this paper, we have proposed a Ricci curvature and Ricci curvature flow framework for directed graphs, extending the geometric analysis tools that have been predominantly applied to undirected networks. A key theory is the establishment of a discrete Ricci curvature flow for strongly connected directed graphs, which guarantees a unique global solution and serves as the foundation for further structural analysis.
Building on this framework, we developed a core subgraph detection algorithm in directed graphs. The method combines Ricci curvature flow evolution with an edge deletion strategy to find the most structurally significant components. Experimental evaluations on real-world networks demonstrate that our approach consistently outperforms classical centrality based methods, achieving superior results on several structural metrics.
Future work may explore other theoretical properties of directed Ricci curvature flows. Additionally, the framework can be extended to directed hypergraphs and dynamic networks, enabling broader applications in complex network analysis.

\section*{Declarations}

\noindent
\textbf{Data availability}:
All data needed are available freely at https://github.com/12tangze12/Finding-core-subgraphs-on-directed-graphs.

\noindent
\textbf{Conflict of interest}: The authors declared no potential conflicts of interest with respect to the research, authorship, and publication of this article.

\noindent
\textbf{Ethics approval}: The research does not involve humans and/or animals. The authors declare that there are no ethics issues to be approved or disclosed.


\begin{thebibliography}{99}

\bibitem{Albert} R. Albert, B. DasGupta, N. Mobasheri, Topological implications of negative curvature for biological and social networks, Phys. Rev. E 89 (2014) 032811.

\bibitem{Bai-Lin} S. Bai, Y. Lin, L. Lu, Z. Wang, S. Yau, Ollivier Ricci-flow on weighted graphs, Amer. J. Math. 146 (2024) 1723-1747.

\bibitem{Bai} S. Bai, R. Li, S. Liu, X. Lai, Ricci flow on weighted digraphs with balancing factor, 	arXiv:2509.19989, 2025.

\bibitem{2011} V. Batagelj,  M. Zaveršnik, An O(m) algorithm for cores decomposition of networks,  Adv. Data Anal. Classif. 5  (2011) 129-145.

\bibitem{Saucan} V. Barkanass, J. Jost, E. Saucan, Geometric sampling of networks, J. Complex Netw. 10 (2022).



\bibitem{1999}  S. P. Borgatti, M. G.  Everett,  Models of core/periphery structures, Social Networks 21 (1999) 375-395.

\bibitem{Eidi} M. Eidi, J. Jost, Ollivier Ricci curvature of directed hypergraphs, Sci. Rep. 10 (2020) 12466.

\bibitem{Eswaran1976} K. P. Eswaran, R. E. Tarjan, 
Augmentation Problems,  SIAM J. Comput. 5 (1976) 653-665. 


\bibitem{ref17} R. Forman, Bochner's method for cell complexes and combinatorial Ricci curvature, Discrete Comput. Geom. 29 (2003) 323-374.


\bibitem{freeman} L. C. Freeman, Centrality in social networks conceptual clarification, Social Networks 1 (1979) 215-239.


\bibitem{hamilton1982ricci} R. Hamilton,
Three-manifolds with positive Ricci curvature,
J. Differ. Geom.  17 (1982)  255-306.

\bibitem{human} J. Kunegis, Konect: the Koblenz network collection, Proceedings of the 22nd International Conference on World Wide Web,   (2013) 1343-1350. 

\bibitem{Kitazono} J. Kitazono, R. Kanai, M. Oizumi, Efficient search for informational cores in complex systems: Application to brain networks, Neural Netw. 132, (2020) 232-244.

\bibitem{Koujaku} S. Koujaku, I. Takigawa, M. Kudo, and H. Imai, Dense core model for cohesive subgraph discovery, Social Networks 44 (2016) 143-152.



\bibitem{Lai1} X. Lai, S. Bai, Y. Lin,
Normalized discrete Ricci flow used in community detection,
Phys. A  597 (2022)  127251.

\bibitem{Li} R. Li, Generalization of Ricci flow on directed graphs and its applications, Bachelor thesis, Renmin University of China, 2024.

\bibitem{Li-Munch} R. W. Li, F. M\"unch, The convergence and uniqueness of a discrete-time nonlinear Markov chain, arXiv: 2407.00314, 2024.

\bibitem{Lin1} Y. Lin, L. Lu, S. T. Yau, Ricci curvature of graphs, Tohoku Math. J. 63 (2011)  605-627.

\bibitem{Ma1} J. Ma, Y. Yang, A modified Ricci flow on arbitrary weighted graph,  J. Geom. Anal. 35 (2025) 332.

\bibitem{Ma2} J. Ma, Y. Yang, Evolution of weights on a connected finite graph, arXiv:2411.06393, 2024.

\bibitem{Ma3} J. Ma, Y. Yang, Piecewise-linear Ricci curvature flows on weighted graphs, arXiv:2505.15\allowbreak395, 2025.


\bibitem{newman2} M. Newman, Networks: An Introduction,
Oxford University Press, 2010.

\bibitem{Ni1} C. C. Ni, Y. Y. Lin, F. Luo, J. Gao,
Community detection on networks with Ricci flow,
Sci. Rep. 9 (2019)  9984.


\bibitem{Ollivier1} Y. Ollivier, Ricci curvature of metric spaces, C. R. Math. 345 (2007) 643-646.

\bibitem{ref25} Y. Ollivier, Ricci curvature of Markov chains on metric spaces, J. Funct. Anal.  256 (2009) 810-864.

\bibitem{Ozawar1} R. Ozawar, Y. Sakurai, T. Yamada, Geometric and spectral properties of directed graphs under a lower Ricci curvature bound, Calc. Var. Partial Differential Equations 59 (2020) 142.

\bibitem{page} L. Page, S. Brin, R. Motwani, T. Winograd,
The PageRank citation ranking: Bringing order to the web,
Technical Report, Stanford Digital Libraries SIDL-WP-1999-0120  (1999) 161-172.


\bibitem{perelman2002entropy} G. Perelman, The entropy formula for the Ricci flow and its geometric applications, arXiv:0211159, 2002.


\bibitem{1983} S. B. Seidman, Network structure and minimum degree, Social Networks  5  (1983) 269-287.


\bibitem{Sengupta} P. Sengupta, N. Azarhooshang, R. Albert, B. DasGupta, Finding influential cores via normalized Ricci flows in directed and undirected hypergraphs with applications, Phys. Rev. E 111 (2025) 044316.



\bibitem{ref18} R. P. Sreejith, K. Mohanraj, J. Jost, E. Saucan, A. Samal, Forman curvature for complex networks,
J. Stat. Mech.  (2016) 063206.


\bibitem{ODE} G. Wang, Z. Zhou, S. Zhu, S. Wang, Ordinary differential equations (in Chinese), Higher Education Press,
 2006.

\bibitem{elegans} D. J. Watts, S. H. Strogatz, Collective dynamics of "small-world" networks. Nature 393,
(1998) 440-442.


\bibitem{Zhao} J. Zhao, J. Ma, Y. Yang, L. Zhao, Core detection via Ricci curvature flows on weighted graphs, arXiv:2508.01400, 2025.







\end{thebibliography}
\end{document}